\documentclass{jfm}
\usepackage{graphicx}
\usepackage{epstopdf, epsfig}

\usepackage{natbib}
\usepackage{graphicx}
\usepackage{float}
\usepackage{subfigure}
\usepackage{amsmath}
\usepackage{amssymb}
\usepackage{color}
\usepackage{soul}
\usepackage{bm}

\graphicspath{{Figures/}}

\newcommand{\ii}[0]{\mathrm{i}}
\newcommand{\dd}[0]{\mathrm{d}}
\newcommand{\ee}[0]{\mathrm{e}}

\shorttitle{Wavepacket modulation in shock-containing jets}
\shortauthor{P. A. S. Nogueira et al.}

\title{Wavepacket modulation in shock-containing jets}

\author{Petr\^onio A. S. Nogueira\aff{1}
  \corresp{\email{petronio.nogueira@monash.edu}},
  Hamish W. A. Self\aff{1},
  Aaron Towne\aff{2},
  \and
  Daniel Edgington-Mitchell\aff{1}}

\affiliation{\aff{1}Department of Mechanical and Aerospace Engineering, Laboratory for Turbulence Research in Aerospace and Combustion, Monash University, Clayton, Australia
\aff{2}Department of Mechanical Engineering, University of Michigan, 2350 Hayward Street, Ann Arbor, MI 48109, USA}

\begin{document}

\maketitle

\begin{abstract}
We propose a new approach to predict the modulation of wavepackets in shock-containing jets. With a modelled ideally expanded mean flow as input, an approximation of the shock-cell structure is obtained from the parabolised stability equations (PSE) at zero frequency. This solution is then used to define a new shock-containing mean flow, which is a function of the shock-cell wavenumber at each streamwise station. Linearisation of the Navier-Stokes equations around this quasi-periodic mean flow allows us to postulate a solution based on the Floquet ansatz, and further manipulation of the equations leads to a system called the parabolised Floquet equations (PFE) that bears several similarities to PSE. The modulation wavenumbers are marched spatially together with the central Kelvin-Helmholtz wavenumber, leading to a modulated wavepacket as the final solution. The limitations of PFE are highlighted, and the method is applied to two sample cases: a canonical slowly diverging jet at low supersonic Mach number and an overexpanded jet, for which large-eddy simulation (LES) data is available. Overall, good agreement is observed between the wavepackets predicted by PFE and the leading spectral proper orthogonal decomposition (SPOD) modes from the LES, suggesting that the method is able to capture the underlying physical mechanism associated with wavepacket modulation: the extraction of energy from the mean flow by the Kelvin-Helmholtz mode and a redistribution of energy to modulation wavenumbers due to the interaction with the shock-cell structure.
\end{abstract}


\section{Introduction}
\label{sec:intro}

Understanding the relevant physical mechanisms in jet noise generation has long been a topic of great interest in the fluids community, with the goal of developing quieter engines. Starting with \cite{mollo1967jet}, attention has been given to large-scale coherent structures in these flows and their connection with the far-field sound. \cite{crow1971orderly} identified these structures, using smoke visualisations, which revealed trains of ``puffs'' with a clear amplitude modulation, comprising regions of growth, saturation and decay. These wavepacket structures have been shown to be responsible for most of the downstream acoustic radiation in turbulent jets \citep{cavalieri2012axisymmetric,jordan2013wave,cavalieri2019amr}. 

The genesis of these coherent structures was found to be in the stability characteristics of the flow: in jets and shear-layers, the presence of an inflection point in the base flow leads to the appearance of an inviscid instability, namely the Kelvin-Helmholtz (KH) mode \citep{michalke1964inviscid,michalke1965spatially}. This mode extracts energy from the mean flow, resulting in the highly amplified structures observed experimentally \citep{crow1971orderly,brown1974density}. This clear instability mechanism inspired the development of several modelling methods to predict the key features of wavepackets. While the application of locally parallel linear stability analysis resulted in only qualitative agreement with experimentally educed KH modes \citep{michalke1971instabilitat,mattingly_chang_1974}, models that consider the jet development were shown to quantitatively capture the overall behaviour of these waves. For instance, \cite{crighton1976stability} proposed a theoretical method based on a WKB approximation to overcome the limitations of the local analysis, obtaining the solution for a slowly-varying mean flow. The parabolised stability equations (PSE), introduced by \cite{bertolotti1991analysis}, improved upon the WKB approach by tracking the slow downstream evolution of the shape function and wavenumber \citep{herbert1994,herbert1997parabolized}. Application of PSE to predict wavepacket shapes and phase velocities was first performed by \cite{gudmundsson2011jfm} for subsonic jets, which led to good agreement with modes extracted from data via spectral proper orthogonal decomposition (SPOD, \cite{towne2018spectral}) for frequencies relevant to jet noise. This comparison has since been extended to higher frequencies and azimuthal wavenumbers \citep{sasaki_2017}, ideally expanded supersonic jets \citep{sinha2014supersonic}, and  jets with azimuthal non-homogeneity \citep{sinha2016parabolized}.

While PSE provides a good approximation of the overall behaviour of wavepackets in turbulent jets at low computational cost, it also has several limitations, such as its inability to predict the wavepacket characteristics at low frequencies \citep{gudmundsson2011jfm}. This problem is addressed if more advanced methods, like resolvent analysis \citep{mckeon2010critical}, are employed. As shown by \cite{towne2018spectral}, response modes predicted by resolvent analysis are closely related to the most energetic structures in the flow educed using SPOD; in fact, if the non-linear terms in the NS equations expanded about the mean flow are uncorrelated in space, SPOD and resolvent response modes are identical. This provides a rigorous explanation for the success of resolvent analysis in predicting coherent structures in various turbulent flows \citep{schmidt2018spectral,lesshafft2019resolvent,pickering2019lift,nogueira2019large,abreu_cavalieri_schlatter_vinuesa_henningson_2020,nogueira_morra_martini_cavalieri_henningson_2021,morra_nogueira_cavalieri_henningson_2021}.

While these modelling approaches are well-suited for the study of subsonic or ideally expanded supersonic jets, their applicability to shock-containing flows is less straightforward. As summarised in recent reviews \citep{RAMAN1998,EdgingtonMitchell2019}, these flows are usually subject to a resonance phenomenon called screech, which leads to the appearance of sharp tones in the acoustic fields of imperfectly expanded jets. As shown recently by \cite{nogueira2021splsa}, this phenomenon is caused by an absolute instability mechanism triggered by the flow periodicity induced by the shock-cells, which causes disturbances to grow in both space and time until they reach a saturated state. This absolute instability leads to unstable global modes, as shown in \cite{BENEDDINE2015} (or marginally stable ones, as in \cite{EdgingtonMitchell2020}). The high computational cost associated with computing global modes in these flows limits their utility for wavepacket modelling. The presence of sharp discontinuities in the flow prohibits the application of PSE and other WKB-based methods as well. As shown by \cite{bertolotti_herbert_spalart_1992}, the PSE approach is only beneficial when the flow properties are slowly varying in the streamwise direction, and when the flow is only convectively unstable (no absolute instability); none of these conditions are satisfied in shock-containing jets. For that reason, most of the broadband shock-associated noise (BBSAN) prediction models are based either on first principles (as in \cite{HARPERBOURNE974,TamTanna1982,TAM1987}) or numerical simulations \citep{morris_miller_2010,tan2019equivalent}. As the key mechanism of this phenomenon is the interaction between instability waves and the shock-cell structure, a low-order model for predicting the equivalent shock-modulated wavepacket for these jets would be useful.

Some low-order alternatives to solve this issue have been proposed in the literature. In \cite{ray_lele_2007}, following the ideas of \cite{TAM1987}, the authors build a BBSAN source term based on the interaction between a shock-cell predicted from linear stability analysis, and a wavepacket obtained using PSE. Even though the final source term may have the expected spectral shape (with energy in wavenumbers associated with the shock-cell structure), since the instability waves are computed using a shock-free mean flow, this modulation may not correspond to that observed in real flows. A similar approach was used by \cite{wong_jordan_maia_cavalieri_kirby_fava_edgington-mitchell_2021}, who obtained the characteristics of the wavepacket using either PSE or SPOD modes from an ideally expanded jet, and the shock-cell structure from experimental data. In both cases, the modulation of the source term of BBSAN (namely, the modulated wavepacket) was obtained \emph{a posteriori}, and the actual effects of the shocks on the KH mode could not be obtained. The periodic, locally parallel formulation developed by \cite{nogueira2021splsa} is able to capture the modulation of this unstable wave under the assumption of spatial periodicity, leading to good agreement with experimental results. Still, the effect of shear-layer spreading and shock-cell variation/weakening cannot be obtained using a periodic model.

In this work, we develop a methodology to explicitly obtain the modulation of wavepackets in shock-containing jets using the parabolised Floquet equations (PFE). This formulation was first developed by \cite{ran2019pfe} for the study of transition mechanisms in boundary layers and is applied for the first time in the present paper to study the coherent structures supported by imperfectly expanded jets. This formulation builds upon the spatially periodic framework developed in \cite{nogueira2021splsa}; here, the effect of the mean flow development can be directly estimated, providing a low-cost method for predicting the overall characteristics of the KH mode in these jets. 

The paper is organised as follows: in section \ref{sec:mathform}, we review the PSE formulation, which will lay the foundations for the present method, and derive the PFE method for the present case. The limitations of the formulation are also highlighted section \ref{sec:mathform}. The results provided by PFE for two sample cases are analysed in sections \ref{sec:modelprob} and \ref{sec:validation}, where comparisons with SPOD modes are also performed. The paper is concluded in section \ref{sec:concl}.

\section{Mathematical formulation}
\label{sec:mathform}

\subsection{Parabolised Stability Equations}
\label{subsec:pse}
We start from the PSE formulation as reviewed by \cite{herbert1997parabolized}, using the Navier-Stokes (N-S) operators described in \cite{towne2016advancements}. In summary, solutions for the compressible linearised N-S equations are sought, which can be written in operator form as

\begin{equation}
    \frac{\partial \mathbf{q}}{\partial t} + \mathbf{A} \mathbf{q} + \mathbf{B} \frac{\partial \mathbf{q}}{\partial x} + \mathbf{C} \frac{\partial \mathbf{q}}{\partial r} + \mathbf{D} \frac{\partial \mathbf{q}}{\partial \theta} + \frac{1}{Re} \mathbf{E}_{r,\theta} \mathbf{q} = 0.
    \label{eqn:NS1}
\end{equation}

The state vector $\mathbf{q}=[\nu \ u \ v \ w \ p]^T$ contains specific volume, streamwise, radial and tangential velocity components and pressure, respectively. $Re$ is the Reynolds number. All spatial coordinates are normalised by the jet diameter $D$, the velocity is normalised by the free-stream sound speed $c_\infty$, and the specific volume is normalised by its value far from the jet ($\nu_\infty$). Following the classical PSE formulation, the second streamwise derivative of the disturbance vector was neglected, and the matrix $\mathbf{E}_{r,\theta}$ is a function of the first and second radial/azimuthal derivatives of the state vector. All the operators are a function of the mean flow $\mathbf{\bar{q}_0}$ and its spatial derivatives. To solve equation (\ref{eqn:NS1}), disturbances are written in the frequency ($\omega$) and azimuthal wavenumber ($m$) domain as

\begin{equation}
    \mathbf{q}(x,r,\theta,t)=\mathbf{\tilde{q}}(x,r,m,\omega) \ee^{- \ii \omega t + \ii m \theta}.
    \label{eqn:NormalModes}
\end{equation}

\noindent The PSE ansatz considers solutions of the form

\begin{equation}
    \mathbf{\tilde{q}}(x,r,m,\omega)=\mathbf{\hat{q}}(x,r,m,\omega) \ee^{\ii \int^x \alpha(x_1) dx_1}=\mathbf{\hat{q}}(x,r,m,\omega) \Gamma(x)
    \label{eqn:solPSE1}
\end{equation}

\noindent with

\begin{equation}
    \Gamma(x)=\ee^{\ii \int^x \alpha(x_1) dx_1},
    \label{eqn:Gamma1}
\end{equation}

\noindent where both $\mathbf{\hat{q}}(x,r,m,\omega)$ and $\alpha(x_1)$ are slowly varying functions of $x,x_1$. This ansatz (and the Leibniz rule) allows us to write the equations as

\begin{equation}
    -\ii \omega \mathbf{\hat{q}} + \mathbf{A} \mathbf{\hat{q}} + \mathbf{B} \frac{\partial \mathbf{\hat{q}}}{\partial x} + \ii \alpha \mathbf{B} \mathbf{\hat{q}} + \mathbf{C} \frac{\partial \mathbf{\hat{q}}}{\partial r} + \ii m \mathbf{D} \mathbf{\hat{q}} + \frac{1}{Re} \mathbf{E}_{r,m} \mathbf{\hat{q}} = 0,
    \label{eqn:NS2}
\end{equation}

\noindent or 

\begin{equation}
    -\ii \omega \mathbf{\hat{q}} + \mathbf{L} \mathbf{\hat{q}} + \mathbf{B} \frac{\partial \mathbf{\hat{q}}}{\partial x} + \ii \alpha \mathbf{B} \mathbf{\hat{q}}= 0,
    \label{eqn:NS3}
\end{equation}

\noindent where $\mathbf{L}=\mathbf{A}+\mathbf{C} \frac{\partial}{\partial r}+ \ii m \mathbf{D}+ \frac{1}{Re} \mathbf{E}_{r,m}$.

In order to remove the ambiguity between the slowly varying variables $\mathbf{\hat{q}}$ and $\alpha$, the normalisation condition

\begin{equation}
    \int_0^{r_{max}}\mathbf{\hat{q}} \frac{\partial \mathbf{\hat{q}}}{\partial x} r \dd r= 0
    \label{eqn:alpha_update}
\end{equation}

\noindent is imposed in the method (see \cite{herbert1994} for further details). By discretizing the streamwise derivatives using first-order implicit Euler differences, (\ref{eqn:NS3}) can be marched in the downstream direction starting from a station close to the nozzle, following the usual PSE method. The method allows us to update the values of $\mathbf{\hat{q}}$ at each axial station by solving the resulting discretized linear system, and the wavenumber $\alpha$ using (\ref{eqn:alpha_update}). This leads to the final solution of the PSE for each frequency and azimuthal wavenumber.

\subsection{Parabolised Floquet Equations}
\label{subsec:pfe}
The modulation of wavepackets caused by the shock-cell structure will be modelled using the parabolised Floquet equations (PFE), first introduced by \citep{ran2019pfe} for the study of interactions between different modes in boundary layer transition. To this point, the entire analysis considered an ideally expanded supersonic mean flow. A first-order approximation \citep{tam_jackson_seiner_1985} of an equivalent shock-containing flow is obtained from the zero-frequency PSE solution as

\begin{equation}
    \mathbf{\bar{q}}=\mathbf{\bar{q}_0}+\mathbf{q_s}\Gamma_s +\mathbf{q_s}^*\Gamma_s^*,
    \label{eqn:ShockMean}
\end{equation}

\noindent where $\mathbf{q_s}$ and $\Gamma_s$ are related to the zero-frequency PSE solution by

\begin{eqnarray}
    \mathbf{q_s}(x,r)=\mathbf{\hat{q}}(x,r,\omega=m=0)\ee^{- \int^x \alpha_i(x_1) dx_1}, \\
    \Gamma_s(x)=\ee^{\ii \int^x \alpha_s(x_1) dx_1},
\end{eqnarray}

\noindent where $\alpha_i$ is the imaginary part of the wavenumber $\alpha$ obtained from the zero-frequency PSE, $\alpha_s$ is the real part, and the superscript $^*$ represents the complex conjugate. Considering that the shock-cells decay quite slowly (around the same rate of the mean flow spreading), $\mathbf{q_s}$ is still a slow-varying function of $x$. This is done such that the term $\Gamma_s$ is associated with the streamwise variation of the shock-cell wavenumber only. Also, since $\alpha_s$ is a real function, $\Gamma_s^*=\ee^{-\ii \int^x \alpha_s(x_1) dx_1}$.

Since the shock-containing mean flow has a component associated with the shock-cell structure (with non-zero streamwise wavenumbers), we may propose a slightly different solution for (\ref{eqn:NS1}), which considers a central wavenumber and the modulation by the shock-cells. This solution can be written as

\begin{equation}
    \mathbf{q}(x,r)=\sum_{j=-N}^N  \mathbf{\hat{q}_j}(x,r) \Gamma(x)\Gamma_s^j(x),
    \label{eqn:solPSE_shocks}
\end{equation}

\noindent where $N$ is the number of harmonics considered in the analysis (the frequency and wavenumber dependence of the solution is implied). In contrast with PSE, our objective is to find not only $\Gamma(x)$ and $\mathbf{\hat{q}_0}(x,r)$, but also the slow-varying parts associated with the modulated wavenumbers $\alpha\pm j\alpha_s$ ($\Gamma(x)\Gamma_s^j(x)$). The outputs of the formulation are the slow-varying parts of the central $\mathbf{\hat{q}_0}$ and modulated $\mathbf{\hat{q}_{\pm j}}$ wavenumbers, and the streamwise development of the central wavenumber $\alpha(x)$; thus, in theory, the solution will account only for the downstream-travelling Kelvin-Helmholtz mode and the modulation related to the shock-cell wavenumber (which is allowed to vary slowly in the streamwise direction). By inserting (\ref{eqn:solPSE_shocks}) into the linearised N-S equations, a set of equations for each modulation component can be written (the detailed process for $N=1$ is shown in appendix \ref{app:linops}), which allows us to write the equivalent PFE system as

\begin{equation}
    -\ii \omega \mathbf{\hat{q}} + \mathbf{L_t} \mathbf{\hat{q}} + \mathbf{B_t} \frac{\partial \mathbf{\hat{q}}}{\partial x} + \ii \alpha \mathbf{B_t} \mathbf{\hat{q}}+ \ii \alpha_s \mathbf{B_{st}} \mathbf{\hat{q}}= 0,
    \label{eqn:ShockPSE}
\end{equation}

\noindent where the vector $\mathbf{\hat{q}}=[\mathbf{\hat{q}_{-N}} \ ... \ \mathbf{\hat{q}_{0}} \ ... \ \mathbf{\hat{q}_{N}}]^T$ comprises all the modulation components. The explicit form of the matrix operators for $N=1$ is shown in appendix \ref{app:linops}.

The above system has exactly the same form as (\ref{eqn:NS3}) and will be solved in the same fashion. The spatial marching is implemented as in PSE, with the wavenumber $\alpha$ being updated in a similar manner by imposing

\begin{equation}
    \int_0^{r_{max}}\mathbf{\hat{q}_0} \frac{\partial \mathbf{\hat{q}_0}}{\partial x} r \dd r= 0.
    \label{eqn:alpha_update_sh}
\end{equation}

Equation (\ref{eqn:alpha_update_sh}) can be viewed as a first approximation of (\ref{eqn:alpha_update}), by considering the amplitude of the modulation components to be small. As in PSE, PFE requires an initial perturbation which will be marched downstream. This solution is obtained by solving the eigenvalue problem defined by

\begin{equation}
    -\ii \omega \mathbf{\hat{q}} + \mathbf{L_t} \mathbf{\hat{q}} + \ii \alpha \mathbf{B_t} \mathbf{\hat{q}}+ \ii \alpha_s \mathbf{B_{st}} \mathbf{\hat{q}}= 0,
    \label{eqn:EigsPFE1}
\end{equation}

\noindent or

\begin{equation}
    \mathbf{F} \mathbf{\hat{q}} = \ii \alpha \mathbf{B_t} \mathbf{\hat{q}},
    \label{eqn:EigsPFE2}
\end{equation}

\noindent which is obtained by neglecting the streamwise derivatives in (\ref{eqn:ShockPSE}), and assuming the mean flow to be spatially periodic. Solving (\ref{eqn:EigsPFE2}) is equivalent to performing a spatially periodic linear stability analysis (SPLSA) \citep{nogueira2021splsa} truncated to the number of harmonics considered in the problem. This means that the initial solution will share the same characteristics from SPLSA, including the periodicity of the solution with regards to the shock-cell wavenumber: if $\alpha_p$ is a complex eigenvalue of the problem, eigenvalues following $\alpha_p \pm n\alpha_s$ (with $n$ an integer) will also be solutions. The resulting eigenfunction will not be locally parallel, but spatially periodic; it may now have energy at modulation wavenumbers $\text{Re}[\alpha_p \pm n\alpha_s]$. This repetition of the spectrum leads to the appearance of upstream modes close to the KH mode in the spectrum, allowing for an interaction between the modes if the shock-cell amplitude is non-zero. This interaction underpins the appearance of screech tones in the flow. For the present case, this interaction is undesirable, as PFE is unable to account for upstream-travelling waves in its original form. The Kelvin-Helmholtz eigenmode is selected as the initial perturbation for all cases studied herein.

\subsection{Step size restriction for PFE}
\label{subsubsec:step}
As shown by \cite{LiMalik1995} and further explored by \cite{towne2019critical}, the presence of stable upstream-travelling waves in the spectrum of PSE may lead to an instability in the spatial march. A stability analysis of the PSE march leads to the minimum step size restriction (or the minimum step one can choose and still avoid instability)

\begin{equation}
    \Delta x > \Delta x_0 = \max \left( - 2 \frac{ \text{Im}[\alpha_u - \alpha_0]}{|\alpha_u-\alpha_0|} \right),
    \label{eqn:deltax_PSE_max}
\end{equation}

\noindent where $\alpha_0$ is the wavenumber of the mode being marched downstream, $\text{Im}$ stands for imaginary part, and the maximum is taken over all upstream-travelling modes $\alpha_u$. Usually, the $\alpha_{u}$ mode that maximises the expression is part of the upstream-travelling free-stream acoustic branch, which for zero free-stream velocity can be approximated as

\begin{equation}
    \alpha_u = - \omega \sqrt{1-z^2},
    \label{eqn:upstream_ac_PSE}
\end{equation}

\noindent where $z \in [0,\infty)$ is equivalent to a transverse wavenumber. Using this expression for $\alpha_u$ and working with the inequality (\ref{eqn:deltax_PSE_max}) leads to the classic PSE step size requirement 

\begin{equation}
    \Delta x_0 = \frac{1}{|\text{Re}[\alpha_0]|}. 
    \label{eqn:deltax_PSE}
\end{equation}

As in PSE, PFE also has requirements regarding the stability of the spatial march. Similar to the spatially periodic analysis \citep{nogueira2021splsa}, the spectrum from PFE is composed of the main PSE spectrum (with wavenumbers $\alpha_{PSE}$) and modes related to the inclusion of the harmonics (wavenumbers $\alpha_{PSE} \pm N \alpha_s$). This leads to the appearance of new upstream acoustic modes in the spectrum with wavenumbers 

\begin{equation}
    \alpha_u = - \omega \sqrt{1-z^2} \pm N \alpha_s.
    \label{eqn:upstream_ac_PFE}
\end{equation}

Thus, the requirement of (\ref{eqn:deltax_PSE_max}) can now be reworked to consider the repetition of the spectrum, which leads to a minimum step size of

\begin{equation}
    \Delta x_0 = \max \left( \frac{1}{|\text{Re}[\alpha_0] \pm N \alpha_s|} \right). 
    \label{eqn:deltax_PFE}
\end{equation}

While the PFE formulation was derived for an incompressible boundary layer in \cite{ran2019pfe}, the authors did not derive the minimum step size of the formulation in that previous work; this is the first time that it is presented. The equation above shows that the minimum step size of PFE can be larger than the one from PSE, depending on the shock-cell wavenumber. This means that the issues from the PSE method are aggravated in PFE, which sets some limits for its applicability. It is expected that PFE would work as well as PSE for low supersonic Mach number in underexpanded jets issuing from purely convergent nozzles (which have high $\alpha_s$); still, as the Mach number is increased, the shock-cell wavenumber decreases for this kind of nozzle, leading to more strict step size requirements. For example, table \ref{tab:deltax0} shows the minimum step size of PFE assuming the march of a Kelvin-Helmholtz (KH) mode with phase velocity $0.7U_j$ ($U_j$ is the ideally expanded jet velocity) for $N=0$, $1$ and $2$ ($St=2\pi\omega D / U_j=0.7$). It is clear that, while $\Delta x_0$ does not change much with $N$ for low supersonic Mach numbers, it can be strongly affected by the presence of the new acoustic branches, depending on the position of the mode in the spectrum and on the shock-cell wavenumber. For some cases, the step-size restriction renders the method impractical, as will be exemplified in the next sections. In order to mitigate the issue, the stabilisation procedure described in \cite{andersson1998stabilization} was also implemented. This relaxes the step-size restriction by replacing the dissipation from the large implicit steps with a damping term whose magnitude is proportional to the minimum step size. One should note that, even though this method addresses one of the issues of the spatial march, it still relies on the knowledge of a minimum step size, and is not expected to solve all the problems with the PSE/PFE formulations \citep{towne2019critical}. \\

\begin{table}
  \begin{center}
\def~{\hphantom{0}}
  \begin{tabular}{cccc}
        $M_j$ & $\Delta x_0$ ($N=0$) & $\Delta x_0$ ($N=1$) & $\Delta x_0$ ($N=2$)  \\[3pt]
        $1.086$ & $0.1592$ & $0.1972$ & $0.1972$  \\ 
        $1.2$ & $0.1592$ & $1.0335$ & $1.0335$  \\ 
        $1.7$ & $0.1592$ & $0.3591$ & $1.4010$
  \end{tabular}
  \caption{Minimum step size of PFE as function of number of harmonics considered and the ideally expanded Mach number for $St=0.7$.}
  \label{tab:deltax0}
  \end{center}
\end{table}

\subsection{Obtaining shock-modulated wavepackets}

With the method outlined, we now proceed to the analysis of the results generated by PFE. The analysis is performed for two sample cases: an $M_j=1.086$ underexpanded jet issuing from a converging nozzle, and a heated $M_j=1.35$ overexpanded jet issuing from a converging-diverging nozzle (design Mach number $M_d=1.5$). The mean flow parameters used in the latter were adjusted to match the conditions from a large eddy simulation (LES). In the first case, the velocity profile proposed by \cite{crighton1976stability} was used to model the spatial development of the jet. A velocity profile based on \cite{sandham2008nonlinear} was used in the second case, which was shown to best model the overall behaviour of the mean streamwise velocity. In both cases, the mean temperature profile is obtained using a Crocco-Busemann approximation. The radial coordinate is discretised using Chebyshev polynomials \citep{weideman2000matlab}, with the radial mapping proposed by \cite{lesshafft2007linear}; the number of radial collocation points was $N_r=200$ for all the cases studied herein. Boundary conditions were implemented as in \cite{lesshafft2007linear}, and the maximum radial position of the domain was chosen as $r_{max}/D=50$.

Spatial linear stability analysis is applied at the initial station of the flow to get the first estimate of the wavenumber and the initial solution to be marched using PSE. The spatial eigenspectrum of the linearised equations for $\omega=m=0$ possess a family of duct-like modes close to the real axis. These eigenvalues have a real part similar to the solution of \cite{pack1950}, and the shapes of the eigenfunctions are also very comparable to this vortex-sheet-like model (especially if the shear layer is very thin, and if the Reynolds number is very high). These eigenvalues are associated with the different shock-cell modes. PSE does a good job in marching this solution further downstream, and it also partially captures the variation in spacing of the shock-cells (see \cite{choi2001prediction}). Still, preliminary results suggest that this variation in spacing may be better captured by using an eddy-viscosity model, as in the cited work, and by the addition of several shock-cell modes. The amplitude and phase of these modes can be obtained using the pressure mismatch at the nozzle plane, as in \cite{tam_jackson_seiner_1985}. Choosing the first shock-cell mode as the initial solution for PSE, the mode is marched from the nozzle ($x=0$) until the desired position downstream. A spatial step of $\Delta x=1/k_{shock}$ was used for the shock-cell computation, where $k_{shock}$ is the first approximation of the shock-cell wavenumber provided by \cite{pack1950}. This step is required to obtain the slowly-varying functions that represent the shock-cell structure, which will be used in PFE.

For PFE, solutions are marched for $N=0$ and $4$, where $N$ is the number of harmonics considered in the problem ($N=0$ is equivalent to PSE); increasing the number of harmonics beyond $N=4$ does not lead to significant changes in the resulting modes. The eigenvalue problem associated with the initial solution of PFE is solved using the Arnoldi method as implemented in the Matlab function \texttt{eigs}. After the PFE march is finished, the resulting slowly-varying functions ($\mathbf{\hat{q}}$ and $\alpha$) are interpolated onto a finer equispaced mesh. The final modulated wavepacket shape is obtained at this step. The step sizes chosen for PSE and PFE are given by (\ref{eqn:deltax_PSE}) and (\ref{eqn:deltax_PFE}), with $\alpha_0=k_{shock}$ (the shock-cell wavenumber from \cite{pack1950}), and $\alpha_0=2\pi St/0.7$ (which considers a KH mode with phase velocity $0.7U_j$) respectively.

\section{Model problem: underexpanded $M_j=1.086$ jet}
\label{sec:modelprob}

\begin{figure}
\centering
\includegraphics[clip=true, trim= 0 0 0 0, width=\textwidth]{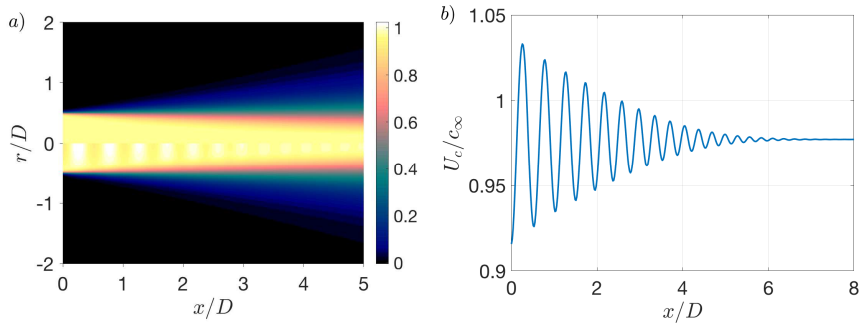}
\caption{Development of the mean streamwise velocity with (bottom) and without (top) the shock-cell structure (a) and mean velocity at the centreline (b) for the shock containing case. The shock-containing solution is obtained by superposing the PSE solution with the ideally expanded mean flow.}
\label{fig:Meanflow21}
\end{figure}

We start with the low-supersonic underexpanded case. As in \cite{crighton1976stability}, the shock-free mean streamwise velocity is modelled as 

\begin{equation}
    U(x,r) = \frac{M}{2}\bigg\{ 1 + \tanh\left[ \frac{a_1}{a_2 x+a_3}\left(\frac{0.5}{r}-\frac{r}{0.5}\right) \right] \bigg\},
    \label{eqn:MeanFlow21}
\end{equation}

\noindent where $M=U_j/c_\infty$ is the acoustic Mach number, and $a_1=10$, $a_2=2.5$, $a_3=1$. Using this equivalent ideally expanded mean flow, the shock-cell structure is computed using PSE, as described in section \ref{subsec:pse}, setting $\Rey=200$. This Reynolds number may appear very low; however, this value functions as the equivalent turbulent Reynolds number for the zero-frequency equation, which is subject to high-amplitude Reynolds stresses. These terms will effectively damp the shock-cells in the streamwise direction, affecting its variation in spacing. This value is also close the one used by \cite{tam_jackson_seiner_1985} in a similar computation. The mean streamwise velocity with and without shock-cells is shown in figure \ref{fig:Meanflow21}(a). For this low-subsonic case, the shock cells are very weak, and most of their effect on the mean flow is in the potential core of the jet; the shear layer is not significantly modified in this case. As expected, the development of the mean flow and the action of viscosity lead to a decrease in the shock-cell amplitude as we go downstream, which is also highlighted in figure \ref{fig:Meanflow21}(b), where the mean streamwise velocity at the centreline is shown. The growth of the shear layer shifts the sonic line towards the centreline, shortening the distance between shock reflection points, which leads to an increase in the shock-cell wavenumber. These effects were already observed in previous works \citep{choi2001prediction,tam_jackson_seiner_1985}, and are highlighted here for clarity.

\begin{figure}
\centering
\includegraphics[clip=true, trim= 0 0 0 0, width=\textwidth]{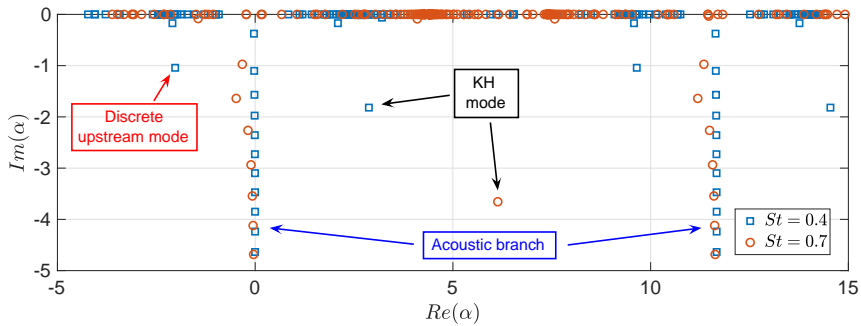}
\caption{Eigenspectra of the truncated spatially periodic linear stability analysis for $St=0.4$ and $0.7$, with $N=4$. Only 500 eigenvalues are shown for each frequency.}
\label{fig:Spec21}
\end{figure}

Following the PFE method, the flow is now linearised around the resulting mean flow obtained from the ideally expanded mean flow and the PSE solution for the shock-cells. In this example case, only axisymmetric disturbances ($m=0$) will be analysed. The initial step of PFE requires the solution of a spatially periodic linear stability analysis truncated to the number of harmonics ($N$) considered in the formulation, where the wavenumber associated with the periodicity is defined by the wavenumber of the shock-cell structure at the initial streamwise station (this is the equivalent of the spatial linear stability analysis for PSE). Figure \ref{fig:Spec21} shows sample eigenspectra for the present flow at $St=0.4$ and $0.7$, where similar features to \cite{nogueira2021splsa} are observed. Firstly, the same modes that appear in the locally parallel case are also evident here; however, these modes now appear several times, with repeated modes offset from each other by $\alpha_s$. This is exemplified by the evanescent acoustic branch: for both frequencies, this branch arises at $Re(\alpha)=0$ and $Re(\alpha)=\alpha_s = 11.6$, with the equivalent occurring for the discrete upstream mode and the KH mode. As mentioned in section \ref{subsubsec:step}, this repetition of the spectrum leads to the stable/evanescent upstream modes being interpreted by the spatial march as unstable downstream-travelling modes. Frequencies where the KH mode is close to the guided jet mode are also expected to be close to the screech frequency, as shown in \cite{nogueira2021splsa}; due to the presence of both downstream- and upstream-travelling waves in the absolute instability phenomenon, the current PFE method is not appropriate to analyse the resulting flow structures associated with resonance arising from absolute instability. Thus, these frequencies will be avoided in the present analysis.

\begin{figure}
\centering
\includegraphics[clip=true, trim= 0 0 0 0, width=\textwidth]{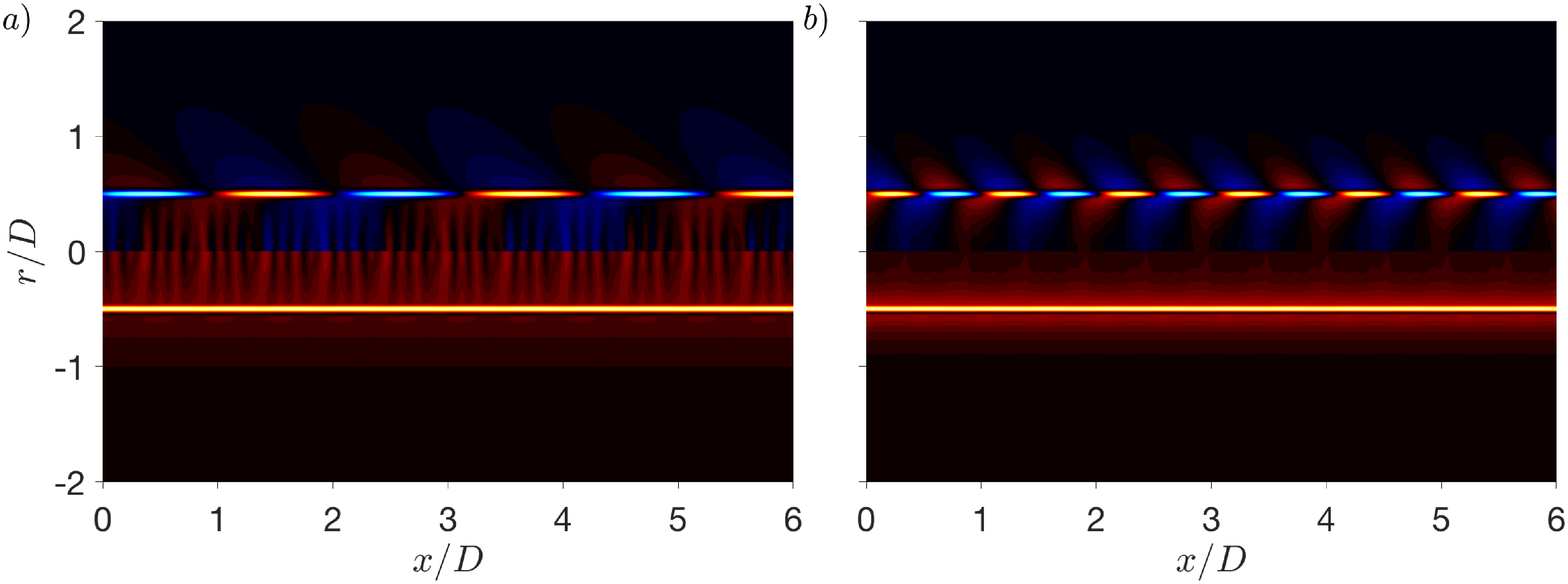}
\includegraphics[clip=true, trim= 0 0 0 0, width=\textwidth]{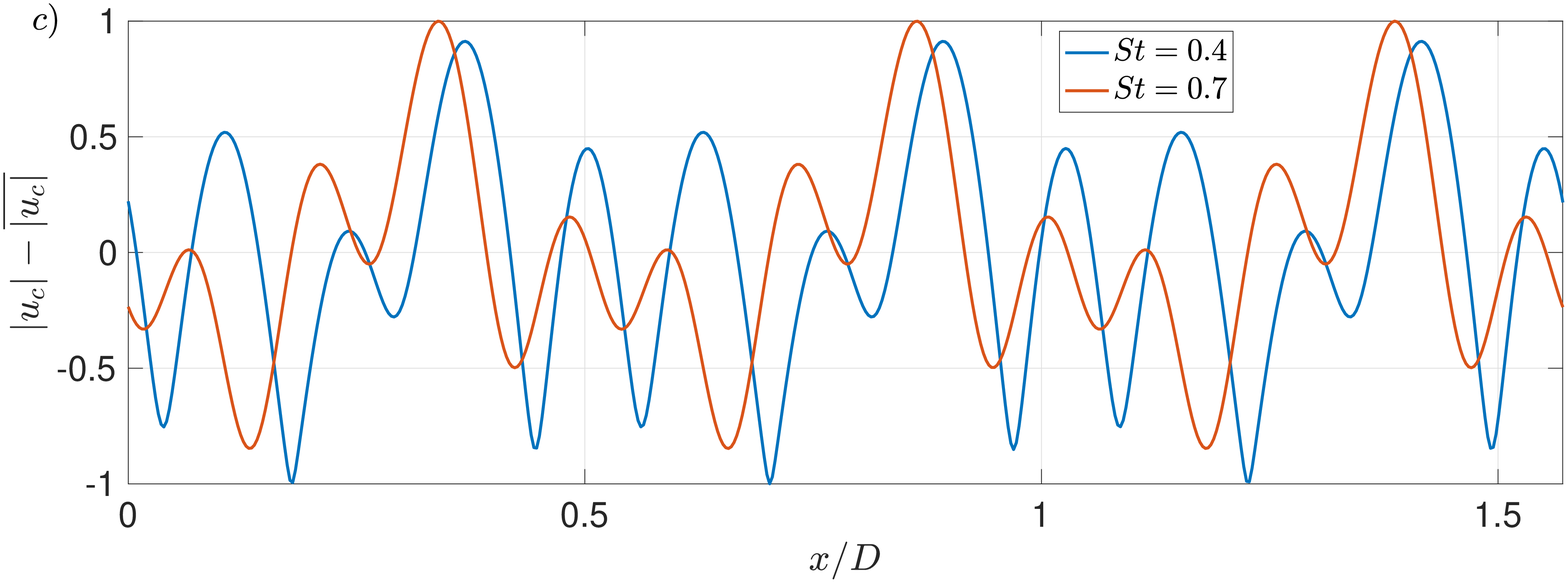}
\caption{Spatial reconstruction of the streamwise velocity fluctuations associated with the initial perturbation (KH) eigenfunctions for $St=0.4$(a) and $0.7$(b); both real part (top) and absolute value (bottom) are shown. Absolute value of the streamwise velocity fluctuations at the centreline subtracted from its streamwise mean (c). In all cases, the imaginary part of $\alpha$ was neglected to highlight the effect of the modulation.}
\label{fig:InitStep}
\end{figure}

Figure \ref{fig:InitStep} shows the shapes of the KH modes considered as the initial solution of PFE for $St=0.4$ and $0.7$. Overall, the shapes of the solutions are similar to the ones presented in \cite{nogueira2021splsa}, with the effects of the modulation observed mainly around the centreline. Comparing figures \ref{fig:InitStep}(a) and (b), it seems that the lower frequency is more affected by modulation components associated with higher harmonics; this is confirmed in figure \ref{fig:InitStep}(c), where the streamwise velocity fluctuations at the centreline are shown. It is also clear that the modulation wavenumbers for both frequencies are exactly the same, which is a consequence of the ansatz used in the formulation. Figure \ref{fig:InitStep} also suggests that, even though there is a modulation for both frequencies, this modulation is fairly weak at the initial step.

\begin{figure}
\centering
\includegraphics[clip=true, trim= 0 60 0 0, width=\textwidth]{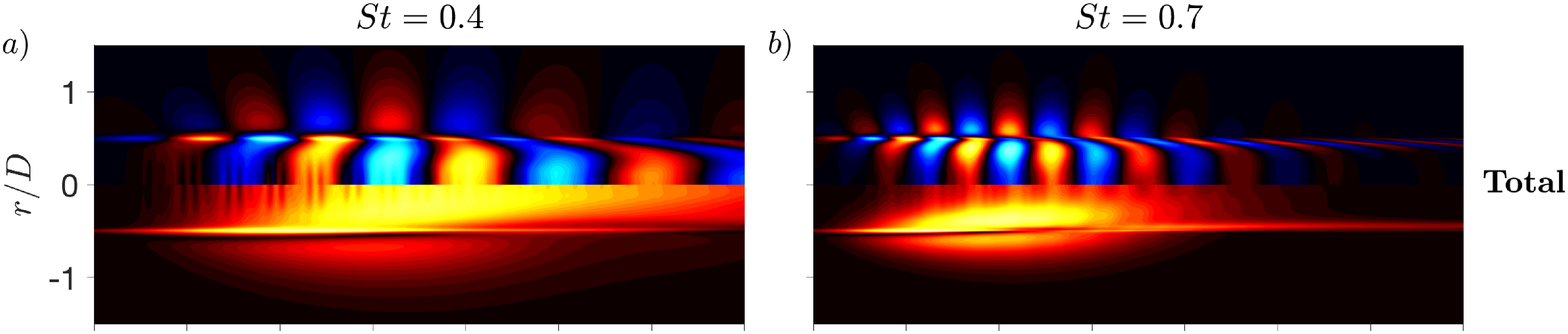}
\includegraphics[clip=true, trim= 0 60 0 30, width=\textwidth]{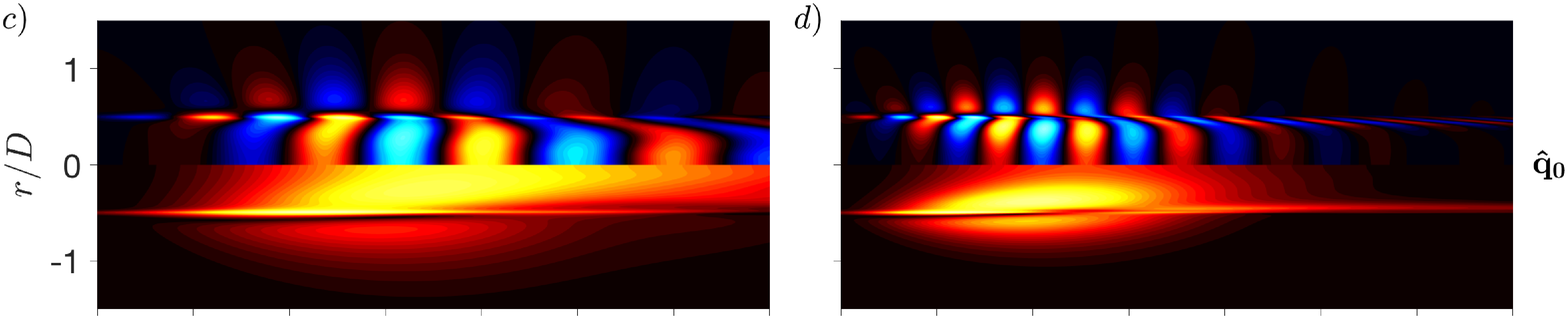}
\includegraphics[clip=true, trim= 0 60 0 30, width=\textwidth]{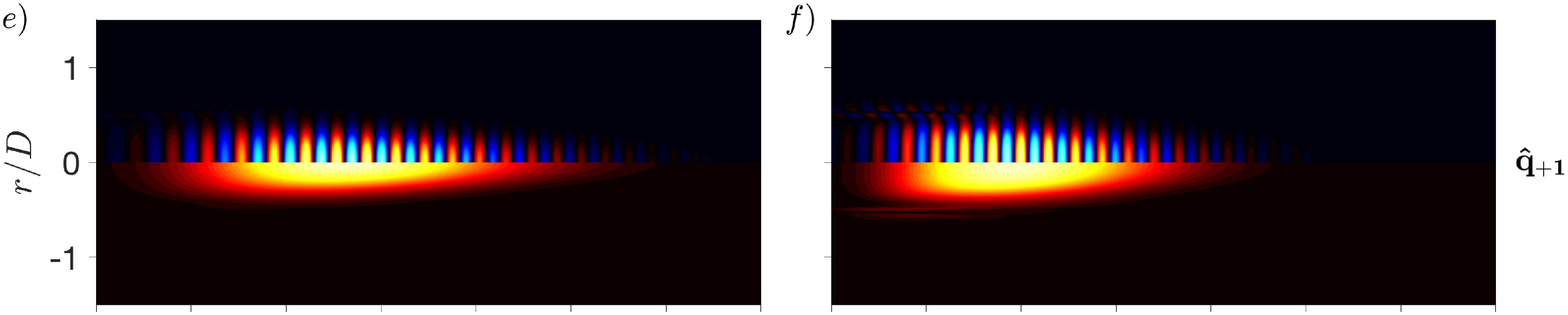}
\includegraphics[clip=true, trim= 0 0 0 30, width=\textwidth]{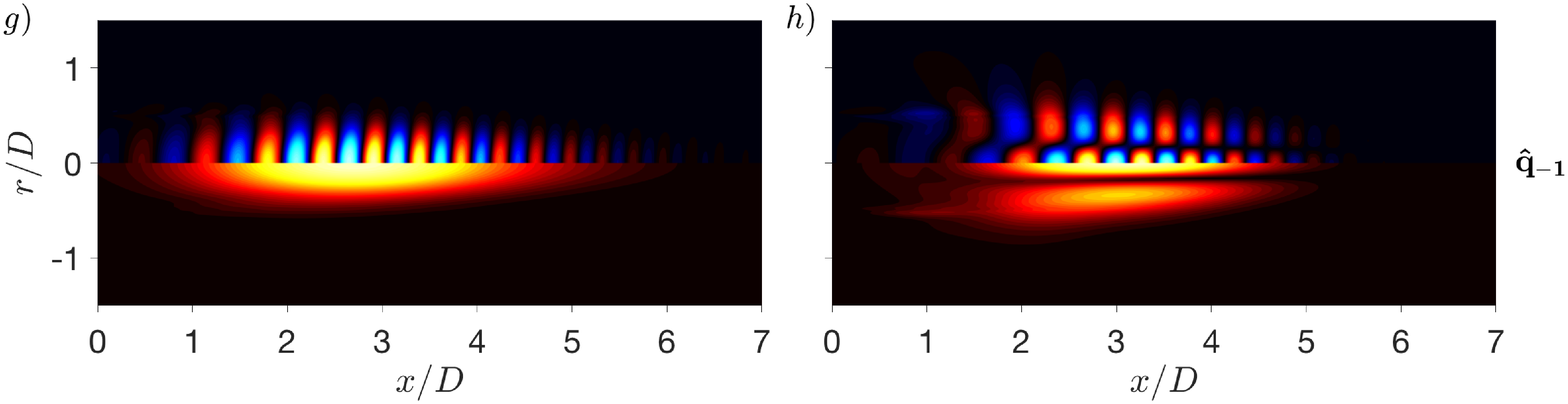}
\caption{Streamwise velocity fields predicted by PFE for $St=0.4$ (a,c,e,g) and $0.7$ (b,d,f,h). Both real part (top) and absolute value (bottom) are shown. Fields are reconstructed in space using all modulation components and $N=4$ in (a,b). Components $\mathbf{\hat{q}_{0}}$ (c,d), $\mathbf{\hat{q}_{+1}}$ (e,f) and $\mathbf{\hat{q}_{-1}}$ (g,h) are also shown.}
\label{fig:NPR21TotVx}
\end{figure}

\begin{figure}
\centering
\includegraphics[clip=true, trim= 0 60 0 0, width=\textwidth]{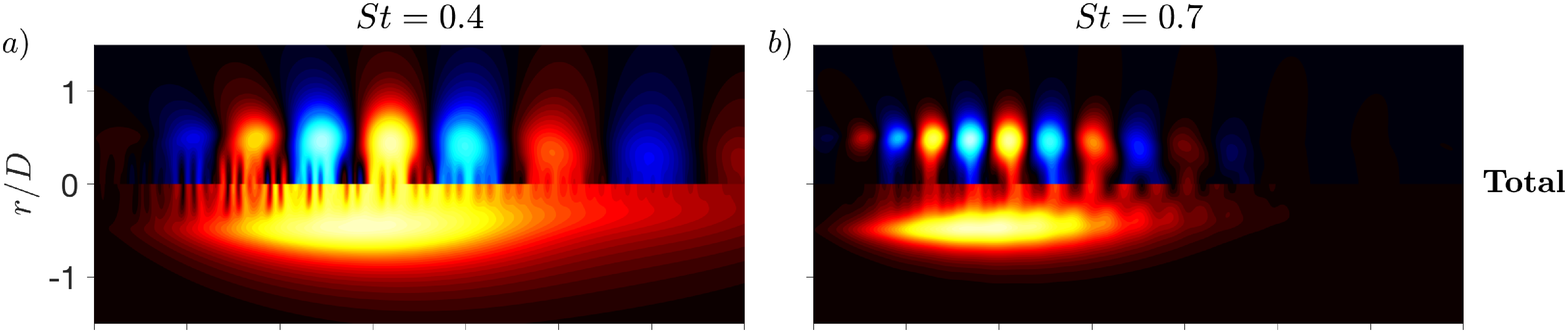}
\includegraphics[clip=true, trim= 0 60 0 30, width=\textwidth]{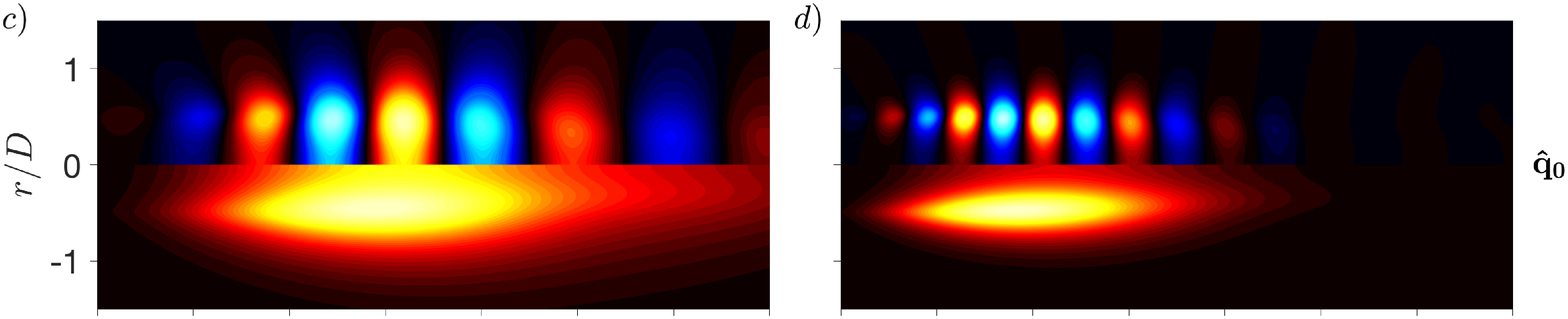}
\includegraphics[clip=true, trim= 0 60 0 30, width=\textwidth]{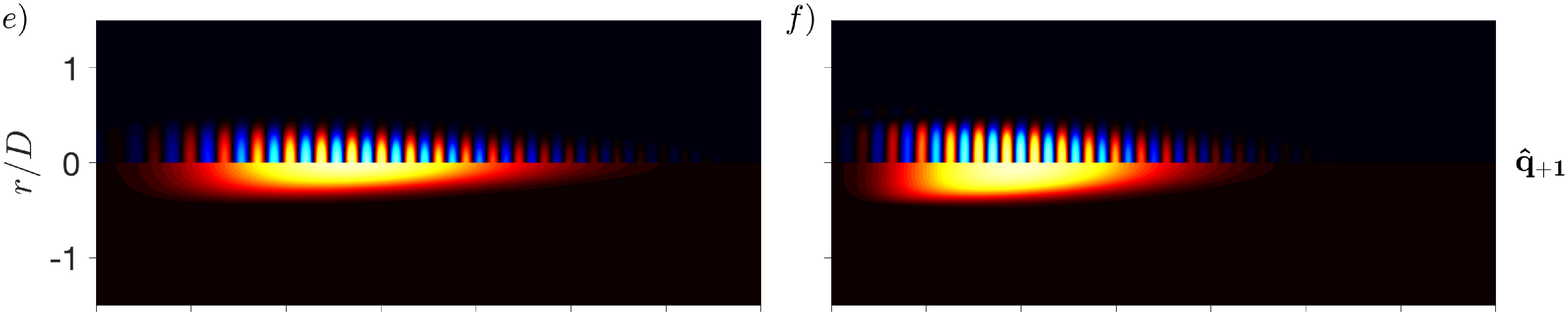}
\includegraphics[clip=true, trim= 0 0 0 30, width=\textwidth]{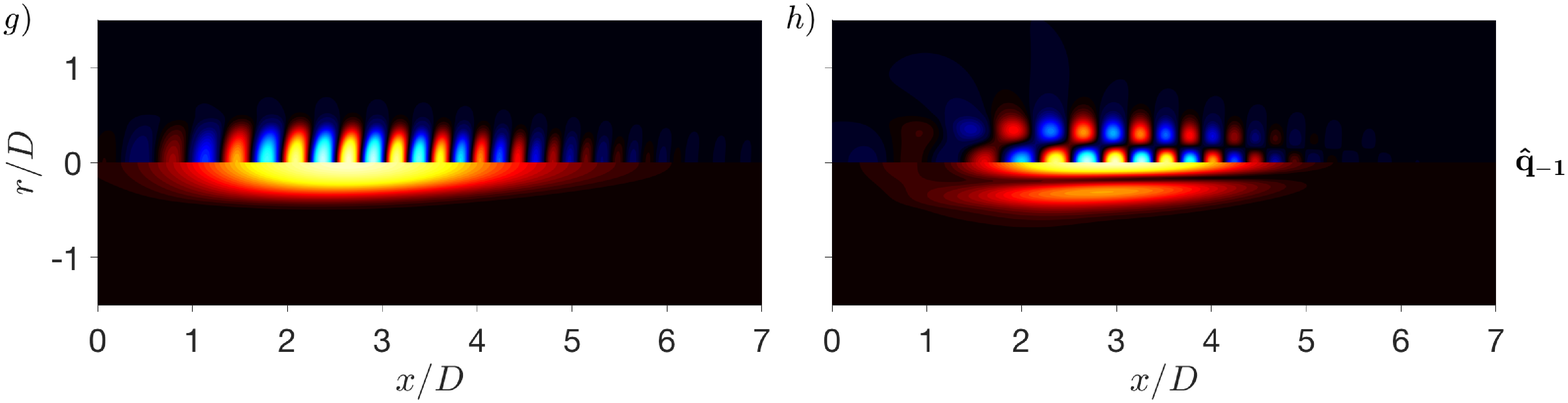}
\caption{Pressure fields predicted by PFE for $St=0.4$ (a,c,e,g) and $0.7$ (b,d,f,h). Both real part (top) and absolute value (bottom) are shown. Fields are reconstructed in space using all modulation components and $N=4$ in (a,b). Components $\mathbf{\hat{q}_{0}}$ (c,d), $\mathbf{\hat{q}_{+1}}$ (e,f) and $\mathbf{\hat{q}_{-1}}$ (g,h) are also shown.}
\label{fig:NPR21TotP}
\end{figure}

The solution shown in figure \ref{fig:InitStep} is marched downstream using the PFE method, which allows us to obtain the modulated wavepackets supported by the shock-containing mean flow at each frequency. Figures \ref{fig:NPR21TotVx} and \ref{fig:NPR21TotP} show, respectively, streamwise velocity and pressure fields obtained using PFE for $St=0.4$ and $0.7$, where both the real part (top) and absolute value (bottom) are shown to highlight the oscillatory behaviour and the modulation of the wavepackets. In these figures, the first row (a,b) depicts the total field obtained from PFE, and the subsequent ones show components $\mathbf{\hat{q}_{0}}$ (c,d), $\mathbf{\hat{q}_{+1}}$ (e,f) and $\mathbf{\hat{q}_{-1}}$ (g,h). Overall, the structures are very similar to previous PSE results (see \cite{sinha2014supersonic}, for instance): the initial KH solution grows exponentially in the early stations of the jet, then it saturates and decays as the shear-layer thickens further downstream. While the oscillatory behaviour and the growth/decay of the solution is similar to the most energetic structures in subsonic and ideally expanded supersonic jets \citep{schmidt2018spectral}, some features of the solutions are unique to PFE, and these are better seen in the absolute value of the reconstructed fields in figures \ref{fig:NPR21TotVx} and \ref{fig:NPR21TotP}. In all cases, the amplitude envelope of the wavepacket has an oscillatory behaviour, especially close to the centreline (where the shock-cell structure has maximum amplitude); these oscillations are associated with the modulation wavenumbers included in the spatial march which, for this case, will lead to the high-wavenumber structures observed in the total field (due to the number of harmonics considered in the analysis).

The wavepacket modulation can be more clearly evaluated by analysing the structure of each modulation component, which can be isolated from the total field directly in (\ref{eqn:solPSE_shocks}); for instance, by neglecting the terms $\mathbf{\hat{q}_{-1}}$ and $\mathbf{\hat{q}_{0}}$, the field associated with the positive modulation ($\mathbf{\hat{q}_{+1}}$) is obtained. The zeroth, first positive and first negative components of the modulation are shown in figures \ref{fig:NPR21TotVx}(c-h) and \ref{fig:NPR21TotP}(c-h). As the modulation of the wavepacket comes from the modulation wavenumbers, the zeroth component is directly comparable to PSE, possessing the same qualitative features. Comparing with the total field, it is also clear that this component is the most energetic one, defining the shape of the final solution. Overall, the shapes of the positive modulation components are very similar for both frequencies, with a single peak close to the centreline and almost zero amplitude in the subsonic region of the jet; this is consistent with the hypothesis that the effect of the shock-cell on the KH mode should be more evident close to the peaks of the shocks in the simplified model. These components also follow the growth/decay behaviour of the KH mode, which agrees with the theory developed by \cite{TamTanna1982}; as the KH mode is the only structure capable of extracting energy from the mean flow, the energy growth of the modulation wavenumbers must be chained to the growth of the unstable mode. As expected, the positive part of the modulation is associated with higher wavenumbers. The negative component of the modulation is also similar for $St=0.4$; as the wavenumber of this structure is given by the difference between the wavenumber of the KH mode and the local wavenumber of the shocks, this modulation leads to structures with larger wavelength (when compared to the positive modulation). Interestingly, the negative part of the modulation for $St=0.7$ (figures \ref{fig:NPR21TotVx}(h) and \ref{fig:NPR21TotP}(h)) differs from its low frequency counterpart, having a double-peak structure. Considering that the second radial order guided jet mode is propagative for this frequency and that the resulting shape of this modulation component is very similar to the cited mode \citep{edgington-mitchell_jaunet_jordan_towne_soria_honnery_2018}, it is possible that the presence of the neutral upstream mode in the spectrum is affecting the spatial march, leading to the appearance of this mode for negative wavenumbers, instead of the modulation of the KH mode. This is also supported by the apparent switch from single- to double-peak structure at the early stations of the jet, also seen in figure \ref{fig:NPR21TotVx}(h); thus, even though the initial solution has a single-peak modulation, the PFE march can induce the appearance of the guided jet mode in the final wavepacket.

This sample case mimics the conditions of the $M_j=1.086$ underexpanded jet analysed in \cite{EdgingtonMitchell2020}. Due to experimental limitations, no time-resolved data could be obtained in the cited work (as is often the case in shock-containing jets), hindering the identification of wavepackets at off-resonance frequencies. In this previous case, proper orthogonal decomposition (POD) was applied to particle-image velocimetry data of the screeching jet, which allowed for the extraction of the most energetic coherent structure associated with the resonance process. This resonance process leads to the appearance of strongly amplified upstream waves, leading to a flow structure whose main features, such as the strong modulation around the shear-layer, are captured by the absolute instability analysis \citep{nogueira2021splsa}. As mentioned earlier, the presence of both upstream- and downstream-travelling waves in the flow at the resonance condition prohibits the application of PFE \citep{bertolotti_herbert_spalart_1992}, and the resulting modulated wavepacket close to the screech frequency can only be analysed as if there was no screech. With that in mind, the structure of the wavepacket for $St=0.7$ also possesses a similar modulation around the centreline, suggesting that some of the features of the modulated wavepacket are preserved in the resonance condition. As mentioned earlier, the structures of the main modulation components (figure \ref{fig:NPR21TotVx}(f,h)) are also very similar to the wave-like structures highlighted in \cite{EdgingtonMitchell2020} for $St=0.7$; this may be due to the presence of both downstream- and upstream-travelling neutral modes in the spatial march, or simply due to the difficulty in separating the structure of the modulation from the different waves supported by the flow.

As mentioned in section \ref{subsubsec:step} (and exemplified in figure \ref{fig:Spec21}), the PFE method is subject to the same issues as PSE, where any stable upstream-travelling mode in the spectrum is considered as an unstable downstream-travelling mode, which may lead to a spurious growth of these modes. This becomes even more serious with the repetition of the spectrum in the PFE method: the spatial step must be chosen to damp the evanescent acoustic branch at $Re(\alpha)=0$ at low frequencies, and also its replica at $Re(\alpha)=\alpha_s$ at higher frequencies. Still, other modes may arise in the analysis that could impact the spatial march. In particular, the neutral upstream acoustic branch may lead to the inclusion of an upstream wave in the march, as shown in the previous section; since this wave is not evanescent, this does not lead to the growth of spurious waves, but may lead to non-physical results due to the parabolisation of the equations. In theory, the correct behaviour of the upstream waves (including their respective modulation) could only be obtained using the equations parabolised in the opposite direction, which would march the desired mode from a given position downstream until the nozzle. This issue with upstream waves becomes worse for frequencies at which the guided jet mode becomes stable \citep{tam_hu_1989,towne_cavalieri_jordan_colonius_schmidt_jaunet_bres_2017,edgington-mitchell_jaunet_jordan_towne_soria_honnery_2018}, which usually occurs at frequencies higher than the screech frequency. In those cases, the spatial step must be severely increased to avoid considering this new mode in the march, which also affects the resulting modulated wavepacket. 

\begin{figure}
\centering
\includegraphics[clip=true, trim= 0 0 0 0, width=\textwidth]{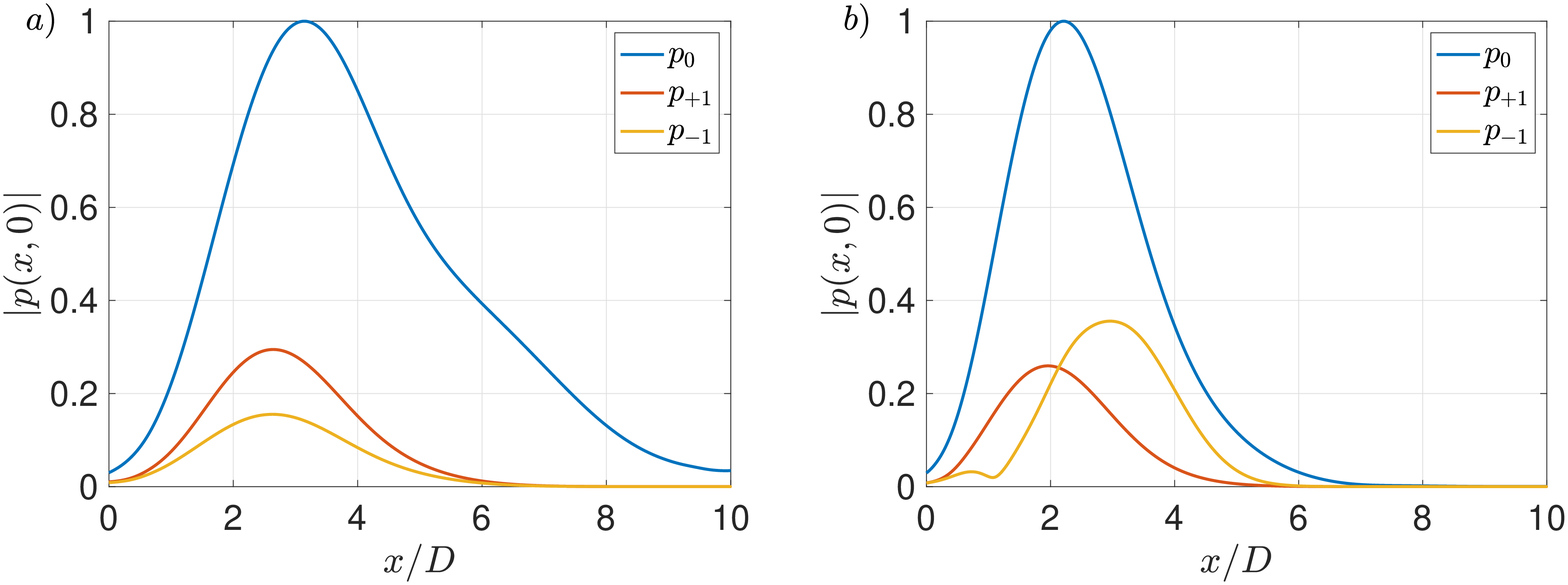}
\includegraphics[clip=true, trim= 0 0 0 0, width=\textwidth]{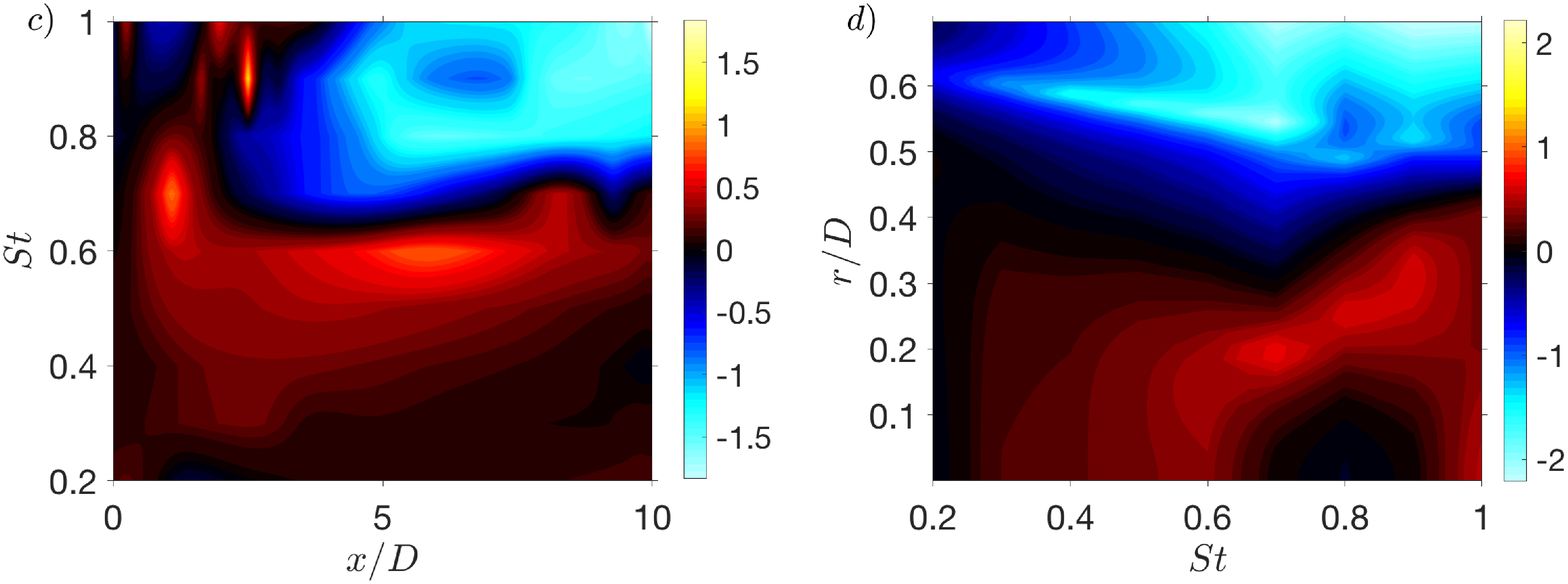}
\caption{Amplitude of the different components for $St=0.4$(a) and $0.7$(b) for $r/D=0$. Ratio between positive and negative components of the modulation ($log(|p_{+1}|/|p_{-1}|)$) for a range of frequencies and $r/D=0$ (c), and $x/D=2$ (d), with $\Delta St=0.1$.}
\label{fig:NPR21_upsmodes}
\end{figure}

This is exemplified in figure \ref{fig:NPR21_upsmodes}. In (a,b), the magnitude of the central solution and the different modulation components at the centreline are shown for $St=0.4$ and $0.7$. For $St=0.4$, the amplitude of the positive modulation $|p_{+1}|$ is higher than the negative one $|p_{-1}|$ at all streamwise stations. That is not the case for $St=0.7$: even though $|p_{+1}|$ is dominant very close to the nozzle, $|p_{-1}|$ increases substantially after reaching low amplitudes; this behaviour is attributed to the presence of the neutral guided jet mode for this frequency. Figure \ref{fig:NPR21_upsmodes}(c) shows the logarithm of the relative magnitude of positive and negative modulation ($log(|p_{+1}|/|p_{-1}|)$) at the centreline for several frequencies and streamwise stations; in this figure, red and blue regions should be interpreted as dominance of the positive and negative modulation components, respectively. This result shows the dominance of the positive modulation up to $St=0.6$, supporting the hypothesis that the neutral guided jet mode (which is propagative for $St>0.67$ at the initial station $x/D=0$) starts to influence the spatial march. For the present case, this wave becomes stable at the initial streamwise station for $St=0.8$; thus, for that frequency, the damping factor of \cite{andersson1998stabilization} had to be further increased to avoid a dominance of the stable guided jet mode. Overall, the radial structure of the negative modulation component is still very similar to the one for $St=0.7$, but its streamwise behaviour is less clear; for these high frequencies, there is no guarantee that the results will correspond to the physics of the problem, especially concerning this component of the modulation. A similar trend is observed in figure \ref{fig:NPR21_upsmodes}(d), where the radial behaviour of the modulation ratio is shown for several frequencies: for low $St$, $|p_{+1}|$ dominates over $|p_{-1}|$ until positions close to the shear layer, suggesting that positive modulation decays faster in the radial direction. For $St \geq 0.7$, a red peak associated with the phase change of the neutral upstream wave is observed around $r/D=0.2$. Again, the negative modulation for these high frequency cases may be contaminated by stable upstream waves, which may change its radial shape.

In summary, these results suggest that one should be cautious when applying PFE for high frequencies, especially when the shock-cell wavenumber is small (as is usually the case in highly underexpanded jets). With that in mind, we can now proceed to a validation case, where the PFE results will be compared to SPOD modes of an overexpanded jet.

\section{Validation: overexpanded $M_j=1.35$ jet}
\label{sec:validation}

Next, we explore the large-eddy simulation database similar to the one described in \cite{bres2017unstructured} (also performed using the compressible flow solver ``Charles'', developed by Cascade Technologies), which comprises an overexpanded jet issuing from a round converging-diverging nozzle (design Mach number $M_d=1.5$) at an ideally expanded jet Mach number $M_j=1.35$ and nozzle temperature ratio $NTR=2.53$. The main differences compared to this previous simulation are the Reynolds number ($\Rey_j=760000$), the presence of wall-modelling inside the nozzle and a near-wall adaptive mesh refinement (see \cite{bres2018jfm}). The total simulation time was $\Delta t c_\infty/D=500$, and fields were saved at a sampling frequency of $fD/c_\infty=10$, resulting in 5000 snapshots. The data are interpolated onto a structured cylindrical grid with dimensions $N_x \times N_r \times N_\theta = 698 \times 136 \times 128$, ranging $0 \leq x/D \leq 30$, $0 \leq r/D \leq 6$, and an azimuthal decomposition is performed using $n_{fft,\theta}=128$. Further details about the LES simulation can be found in \cite{bres2017unstructured}.

As in \cite{gudmundsson2011jfm,cavalieri2013wavepackets,sinha2014supersonic,sasaki_2017,schmidt2018spectral}, PFE results will be compared to the most energetic structures extracted from the flow using spectral proper orthogonal decomposition (SPOD). Following \cite{towne2018spectral}, these structures are obtained by solving the integral eigenvalue problem

\begin{equation}
\int\displaylimits_\Omega \mathsfbi{S}(x,x',r,r',m,\omega) \bm{\psi} (x',r',m,\omega) r'dr'dx' = \sigma (m,\omega) \bm{\psi} (x,r,m,\omega),
\label{eqn:PODint}
\end{equation}

\noindent where $\mathsfbi{S}$ is the two-point cross-spectral density (CSD) tensor of the flow fluctuations (which may be built using any of the available variables), and $\Omega$ is the computational domain. Solution of (\ref{eqn:PODint}) leads to real, positive eigenvalues $\sigma$ (which are associated with the relative energy of each mode), and orthogonal eigenfunctions $\bm{\psi}$. Thus, the most energetic structures of the jet are associated with the first eigenvalue $\sigma_1$, and its respective eigenfunction $\bm{\psi}_1$ for each frequency. The CSD tensor is computed for a reduced domain ($0 \leq x/D \leq 10$, $0 \leq r/D \leq 2$) using $n_{fft}=512$ and $75\%$ overlap, leading to 36 blocks in the Welch method, and a discretisation in frequency of $\Delta St \approx 0.01$. In the present case, we will focus on helical ($m=1$) pressure disturbances. 

\begin{figure}
\centering
\includegraphics[clip=true, trim= 0 0 0 0, width=\textwidth]{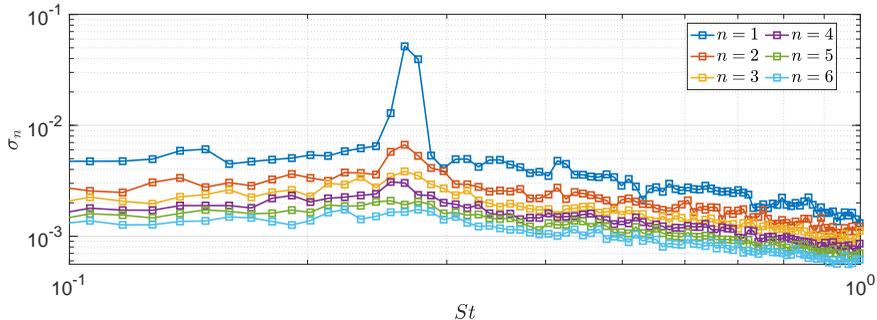}
\caption{First 6 spectral proper orthogonal decomposition eigenvalues as function of Strouhal number.}
\label{fig:SPODGains}
\end{figure}

The SPOD eigen-spectrum is shown in figure \ref{fig:SPODGains}. Unlike subsonic and ideally expanded supersonic jets \citep{schmidt2018spectral}, the spectrum displays a sharp peak around $St=0.26$, associated with the screech tone. This tone is a direct consequence of an absolute instability mechanism induced by the shock-cell periodicity, where flow disturbances grow in both downstream and upstream directions \citep{nogueira2021splsa}. Since PFE can only obtain downstream-travelling waves, the flow structure associated with this tone will not be further analysed herein.

Following the same methodology described in section \ref{sec:mathform}, we define a realistic flow model as an input for both PSE and PFE. For this case, the equivalent ideally expanded streamwise velocity is based on the expression proposed by \cite{sandham2008nonlinear}, which is given by

\begin{equation}
    U(x,r) = \frac{M}{2}\left[ \tanh \left( \frac{r+a(x)}{\delta(x)} \right) - \tanh \left( \frac{r-a(x)}{\delta(x)} \right) \right],
    \label{eqn:MeanFlow135}
\end{equation}

\noindent where

\begin{equation}
    a(x) = D_j \left[ 0.59+0.09\tanh\left(\sqrt{x/\kappa} -2.9 \right) \right]
    \label{eqn:MeanFlow135_a}
\end{equation}

\noindent and

\begin{equation}
    \delta(x) = \frac{7+70(x/\kappa)+0.15(x/\kappa)^4}{1000+(x/\kappa)^3}.
    \label{eqn:MeanFlow135_delta}
\end{equation}

The coefficients in (\ref{eqn:MeanFlow135_delta}) were chosen to provide a similar development of the sonic line when compared to the LES (as zero-frequency PSE results are very sensitive to that parameter, especially concerning the variation in shock-cell spacing further downstream), and $\kappa=1.475$ was obtained by means of a least-squares-based fit of the mean streamwise velocity at the centreline. The ideally expanded diameter $D_j$ (normalised by $D$) is obtained as in \cite{tam_jackson_seiner_1985}. As in the previous case, the temperature and density profiles were obtained using a Crocco-Busemann approximation and the ideal gas law. The mean radial velocity was also neglected, and the Reynolds number for the PSE march was defined as $\Rey=175$ (which is considered to be equivalent to a turbulent Reynolds number for the mean flow equation in the method).

\begin{figure}
\centering
\includegraphics[clip=true, trim= 0 0 0 0, width=\textwidth]{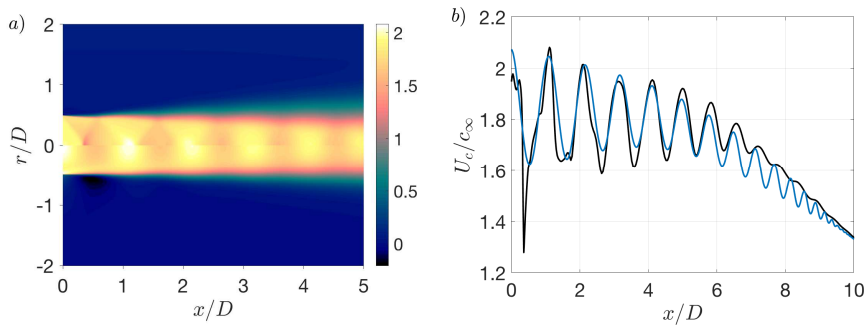}
\caption{Mean streamwise velocity computed from the LES (top) and by a superposition of a modelled ideally expanded flow and the shock-cell structure predicted using zero-frequency PSE (bottom) (a). Comparison between the mean streamwise velocity at the centreline from LES and the present method (b).}
\label{fig:MeanPSELES}
\end{figure}

The resulting mean streamwise velocity predicted from zero-frequency PSE is compared to the LES results in figure \ref{fig:MeanPSELES}. As shown in figure \ref{fig:MeanPSELES}(a), the overall behaviour of the mean flow is well captured by the model, especially concerning the shock-cells. The shear-layer from the LES is thicker at some streamwise stations, but the sonic lines for the two cases are comparable, which led to a good agreement for the shock-cells further downstream. The PSE is unable to account for the Mach reflection near the centreline of the first shock cell, resulting in an overprediction of velocity, but both the amplitude and spacing of subsequent shock cells is predicted well, in line with the results of \cite{HARPERBOURNE974}. This is confirmed in figure \ref{fig:MeanPSELES}(b), where the mean streamwise velocities at the centreline are compared. PSE seems to capture the general behaviour of the shock-cell, including its amplitude, until about $x/D=6$; after that, this structure is strongly damped in the LES, an effect that is not captured by the model. Still, considering the differences in the mean flow characteristics, and that only a single shock-cell mode is included in the analysis, these results can be considered as a good first approximation of the shock-cell train in this turbulent jet. Better agreement with the simulation data (including the sharper changes associated with the shocks) may be obtained if more modes are included in the analysis \citep{choi2001prediction,tam_jackson_seiner_1985}. For the expansion conditions of the present jet, \cite{pack1950} predicts a shock-cell wavenumber of $k_{shock}D=5.3$, which is significantly smaller than the case studied in the previous section. Thus, it is expected that PFE may struggle to produce good predictions even at mid-high frequencies, since the range of frequencies where the KH mode is far from any upstream mode in the PFE eigenspectrum is rather small.

\begin{figure}
\centering
\includegraphics[clip=true, trim= 0 0 0 0, width=\textwidth]{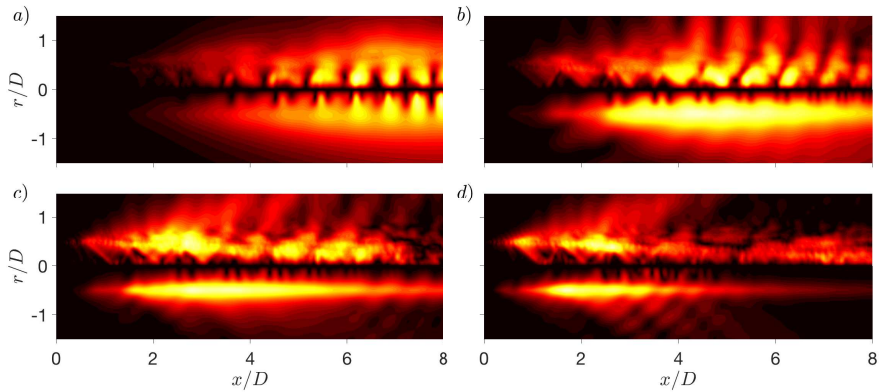}
\caption{Comparison between leading SPOD modes from LES (top) and the flow structure predicted by PFE (bottom) using $N=4$ and several frequencies. Absolute value of the modes are shown for $St=0.19$(a), $0.4$(b), $0.74$(c) and $0.91$(d).}
\label{fig:PFELEScompabs}
\end{figure}

\begin{figure}
\centering
\includegraphics[clip=true, trim= 0 0 0 0, width=\textwidth]{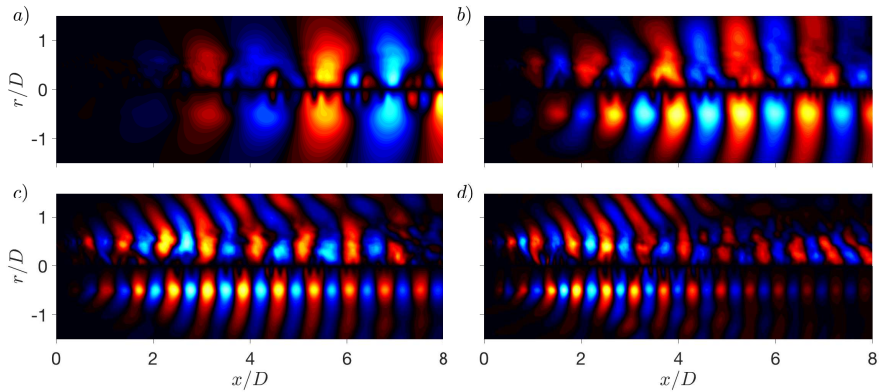}
\caption{Comparison between leading SPOD modes from LES (top) and the flow structure predicted by PFE (bottom) using $N=4$ and several frequencies. Real part of the modes are shown for $St=0.19$(a), $0.4$(b), $0.74$(c) and $0.91$(d).}
\label{fig:PFELEScompreal}
\end{figure}

A comparison between the first SPOD mode from LES and the wavepackets predicted using PFE is shown in figures \ref{fig:PFELEScompabs} (absolute value) and \ref{fig:PFELEScompreal} (real part), for four sample frequencies ($St=0.19$, $0.4$, $0.74$ and $0.91$, chosen as to avoid the cases with very high spatial steps). For this case, given the step-size restrictions and the repetition of the spectrum, the lowest frequency ($St=0.19$) is one of the most favourable case analysed here, since no upstream waves are expected to strongly affect the spatial march. For this case, the PFE seems to capture the overall behaviour of the amplitude envelope, which extends until the end of the domain. While the modes do not seem to capture the modulation particularly well at the early stations of the jet, it performs very well closer to the peak of the KH mode, especially near the centreline (where the shocks are strongest). The three other cases are for frequencies higher than the screech tone, and will be potentially affected by upstream modes in the spectrum. Nevertheless, the overall envelope length is very similar for all cases, and PFE also manages to capture some key features of the modulation. For $St=0.4$, the SPOD mode has a clear standing wave pattern, suggesting the presence of strong upstream waves in the flow; obviously, PFE is not able to capture these waves, and the modulation around the shear-layer is under-predicted. A similar behaviour is observed for $St=0.74$, where a modulation due to upstream acoustic waves is also observed in the outer part of the shear-layer. For $St=0.91$, multiple acoustic beams are identified in both SPOD and PFE (better seen in figure \ref{fig:PFELEScompreal}), with the latter being modulated by the presence of vertical waves in the flow (again, due to the proximity of the KH mode to the evanescent acoustic branch). These acoustic waves are damped in PFE due to the stabilisation methods intrinsic to the formulation, as explained in \cite{towne2019critical}. As in previous cases \citep{sasaki_2017}, a second structure is observed further downstream for the highest frequency in the SPOD, which is not captured by the present model.

The real part of both SPOD and model results are shown in figure \ref{fig:PFELEScompreal}, where the wavelengths of the structures can be compared. As mentioned in \cite{sinha2014supersonic}, PSE can predict moderately different wavelengths due to its normalisation condition, and this is also shown to occur for PFE (which relies on the same condition). This comparison is harder to perform at low frequencies, where the modulation is stronger and wavenumbers of higher magnitude have large amounts of energy; for those cases, this comparison must be performed in the outer region of the jet, where the flow is dominated by the central wavenumbers of the KH mode ($\mathbf{\hat{q}_{0}}$). The radial structures of both the model and the SPOD modes are also in good agreement (similar to PSE). Acoustic waves are more clearly observed in the near field of the jet for $St=0.74$ and $0.91$, but these are poorly predicted by the PFE, as highlighted earlier. Thus, while the present analysis may be able to predict the modulation characteristics of the hydrodynamic field, the method may not be suitable for the analysis of the near acoustic field. Due to the absence of upstream modes close to the KH mode, PSE may actually perform better in the direct prediction of the downstream noise generated by Mach wave radiation. Some sample results of PSE using the shock-containing mean flow from the LES as input are shown in appendix \ref{app:PSEshock}.

\begin{figure}
\centering
\includegraphics[clip=true, trim= 0 0 0 0, width=\textwidth]{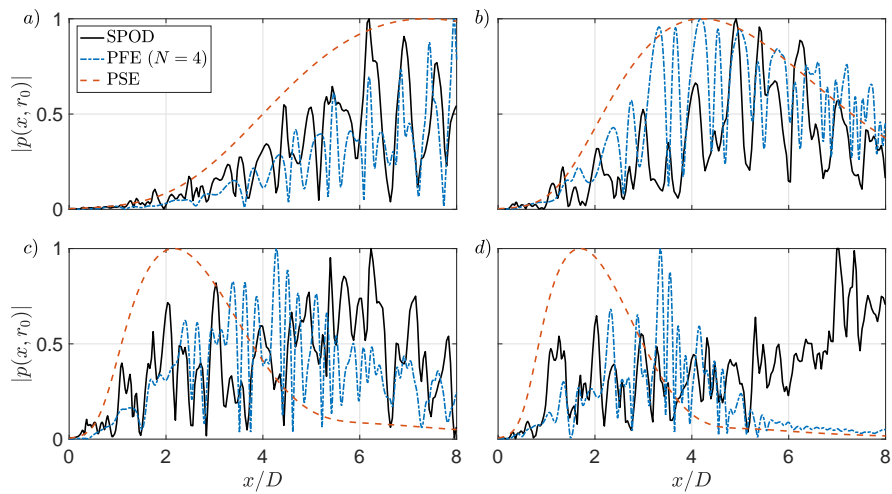}
\caption{Absolute value of leading SPOD modes from LES compared to PFE ($N=4$) and PSE results for several frequencies. Magnitude of the modes extracted at $r/D=0.1$ for $St=0.19$(a), $0.4$(b), $0.74$(c) and $0.91$(d). All curves are normalised by their maximum.}
\label{fig:PFELEScompr01}
\end{figure}

Figure \ref{fig:PFELEScompabs} suggests that, in most cases, the PFE manages to reasonably approximate the modulation closer to the centreline. This is confirmed in figure \ref{fig:PFELEScompr01}, where the absolute value of the modes predicted from PFE are compared to SPOD modes at $r/D=0.1$. Wavepacket shapes predicted by PSE computations using the same ideally expanded mean flow (without shocks) are also plotted for reference. All curves are scaled by their maximum, as the modulation renders the usual scaling at early stations of the jet inappropriate for this case. The positions of the peaks in the modulated wavepackets are in fair agreement with the SPOD in most cases. For lower frequencies, these peaks are either in the same position (as in most streamwise stations for $St=0.19$) or shifted slightly downstream. This may be due to the difficulty of PSE (and PFE) in predicting the exact wavenumbers of the structures; since the modulation is obtained in the model by the constructive/destructive interference of the wave structures related to the different harmonics, a slight shift in the wavenumbers of these structures will lead to a mismatch in the position of the peaks in the reconstructed wavepacket. The prediction of these peaks is slightly improved for higher frequencies, as shown in figures \ref{fig:PFELEScompabs}(c,d), where the bulk of the modulation seems to be captured. Overall, PFE outperforms PSE in the prediction of the envelope of these coherent structures, especially for higher frequencies. For those cases, PSE predicts wavepackets with a faster downstream decay. As PFE considers an energy flux between different wavenumbers (related to the shock-cell), it is possible that this interaction mitigates the decay of the KH mode for those high frequencies. Considering the good agreement between PSE and SPOD modes for ideally expanded jets shown in \cite{sinha2014supersonic}, this also suggests that the presence of a shock-cell structure is the cause for the increase of the wavepacket length for this case, and that PFE is able to capture such a mechanism.

\begin{figure}
\centering
\includegraphics[clip=true, trim= 0 0 0 0, width=\textwidth]{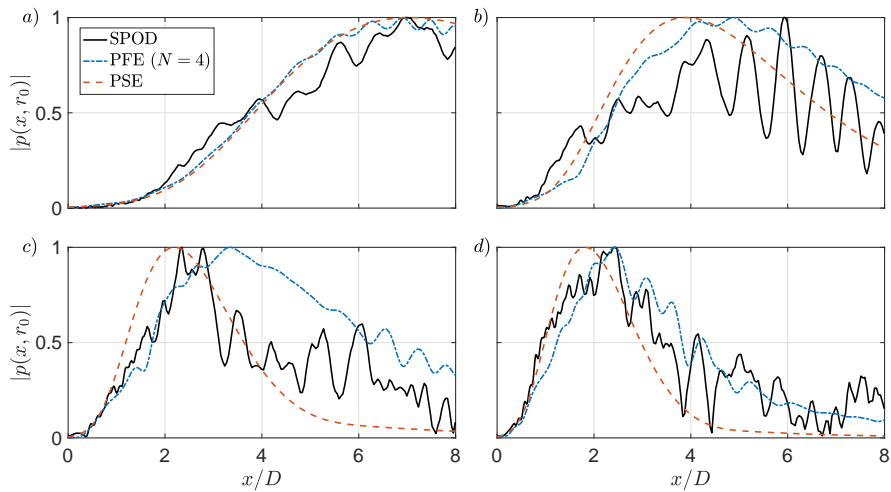}
\caption{Absolute value of leading SPOD modes from LES compared to PFE ($N=4$) and PSE results for several frequencies. Magnitude of the modes extracted at $r/D=0.7$ for $St=0.19$(a), $0.4$(b), $0.74$(c) and $0.91$(d). All curves are normalised by their maximum.}
\label{fig:PFELEScompr07}
\end{figure}

Since the model struggles to predict the near acoustic field, the agreement for higher radial positions deteriorates, especially for higher frequencies (where strong Mach-wave radiation is observed). This is shown in figure \ref{fig:PFELEScompr07}, where $r/D=0.7$ is chosen. The model seems to capture the overall growth of the wavepacket in a similar fashion to PSE for low frequencies, with a small modulation effect. In particular, the standing wave pattern is absent from the model for $St=0.4$ for reasons already highlighted. The modulation effects are more strongly captured for the higher frequencies, where peaks in the growth of the structure are comparable to SPOD modes. A very good agreement for $St=0.91$ is observed at this position, despite the inability of the model to correctly capture acoustic waves.

\begin{figure}
\centering
\includegraphics[clip=true, trim= 0 0 0 0, width=\textwidth]{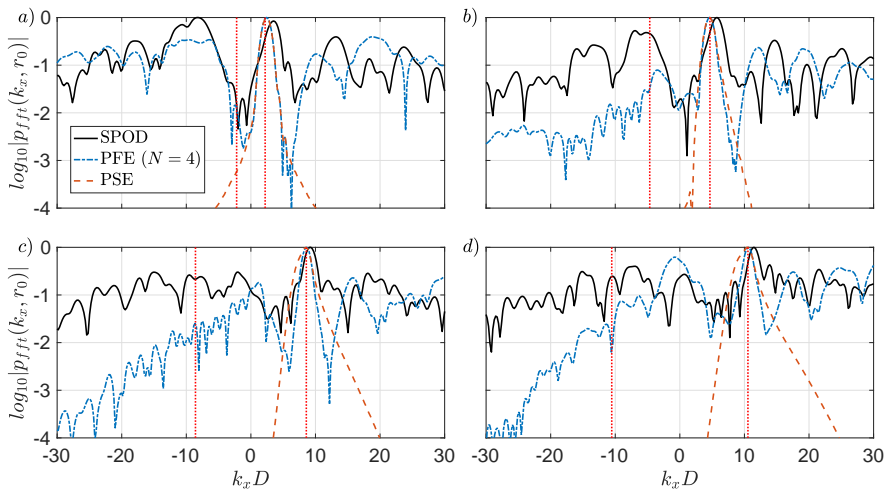}
\caption{Logarithm of the absolute value of leading SPOD modes from LES compared to PFE ($N=4$) and PSE results for several frequencies in the wavenumber domain. Magnitude of the modes extracted at $r/D=0.1$ for $St=0.19$(a), $0.4$(b), $0.74$(c) and $0.91$(d). Dotted lines indicate the acoustic wavenumbers. All curves are normalised by their maximum.}
\label{fig:PFELESwaven}
\end{figure}

To evaluate the ability of PFE to capture the spectral energy content of the resulting wavepackets, a comparison with SPOD modes in the wavenumber domain is shown in figure \ref{fig:PFELESwaven}. Since results were generated for $x/D \leq 10$, the spatial Fourier transforms may be contaminated by domain truncation; a Hann window was applied to both SPOD and PFE modes to minimise these effects. For all frequencies, PSE performs reasonably well in capturing the peak wavenumber associated with the wavepackets in this flow, but other energetic wavenumbers observed in the SPOD modes are suppressed due to the PSE regularisation \citep{towne2019critical}. The energy of the PSE results at negative wavenumbers (responsible for upstream noise radiation) is several orders of magnitude lower than the peak for low $St$, and almost negligible for higher frequencies. While not perfectly aligned, the PFE wavenumber spectra bear several similarities with the SPOD spectra, including the presence of peaks at similar wavenumbers. The absence of upstream waves in the PFE leads to an overall less energetic spectrum for negative wavenumbers, especially for higher frequencies. Still, both PFE and SPOD have a strong energy content at negative propagative wavenumbers (indicated by the dotted lines in figure \ref{fig:PFELESwaven}), suggesting that propagation of PFE wavepackets to the far-field using an acoustic analogy \citep{lighthill1952sga} may lead to the correct directivity \citep{jordan2013wave} and that BBSAN is generated by wavepacket-shock modulation.

Overall, the results presented in this section showed that, despite its limitations, PFE may be a good prediction tool for shock-containing jets. One should be aware of all the modelling steps taken to obtain these results: \emph{i)} a consideration of a mean flow that is not exactly what is observed in the LES; \emph{ii)} the approximation of a train of shock-cells (which are related to steep variations of fluid properties) as a duct-like mode predicted using a linear stability model; \emph{iii)} the truncation of the solution to a limited number of harmonics; \emph{iv)} the absence of turbulent forcing/viscosity in the flow, with coherent structures being predicted using a linear model; \emph{v)} the application of the model to a case where it is not supposed to perform ideally. With all those issues in mind, it is remarkable that so many features of the modulated wavepackets are captured by such a severely simplified model. As in PSE, the PFE formulation does not need any input from experiments, except for an equivalent ideally expanded flow (which can be modelled as in the present analysis), which makes the method suitable for the prediction of coherent structures where no time-resolved data is available, as is the case in the vast majority of shock-containing jets. 

In order to provide a more appropriate comparison, SPOD modes are next filtered to consider only positive wavenumbers (using a simple rectangular bandpass filter), and PFE modes were reconstructed using only their non-negative components (in this case, $\hat{q}_{0}$ and $\hat{q}_{+n}$, with $n=1, 2, 3, 4$). One should keep in mind that, even though the negative wavenumbers are not considered in this comparison directly, their effect on the central and modulation wavenumbers will still be present; for instance, $\hat{q}_{0}$ is still affected by $\hat{q}_{-1}$ in the solution of the equations as all components are marched together. This will only minimise the errors related to the prediction of the negative part of the modulation in the reconstructed mode.  

\begin{figure}
\centering
\includegraphics[clip=true, trim= 0 0 0 0, width=\textwidth]{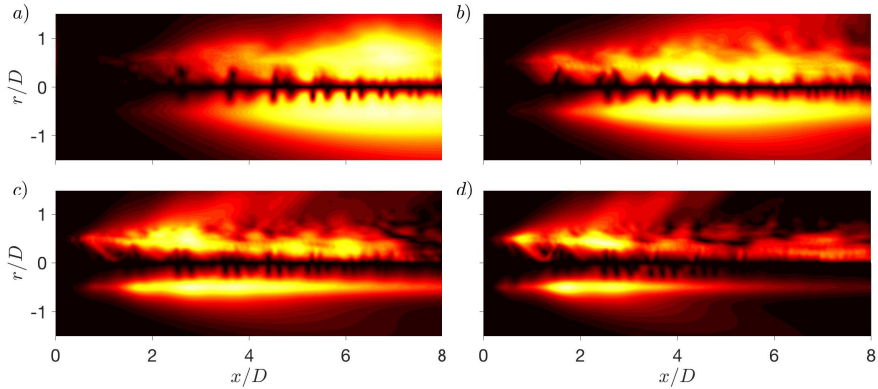}
\caption{Comparison between leading SPOD modes from LES (top) and the flow structure predicted by PFE (bottom) using $N=4$ and several frequencies. Absolute value of the modes are shown for $St=0.19$(a), $0.4$(b), $0.74$(c) and $0.91$(d). Only positive wavenumbers are allowed in the SPOD, and the PFE reconstruction is performed using non-negative components.}
\label{fig:PFELEScompabs_filt}
\end{figure}

\begin{figure}
\centering
\includegraphics[clip=true, trim= 0 0 0 0, width=\textwidth]{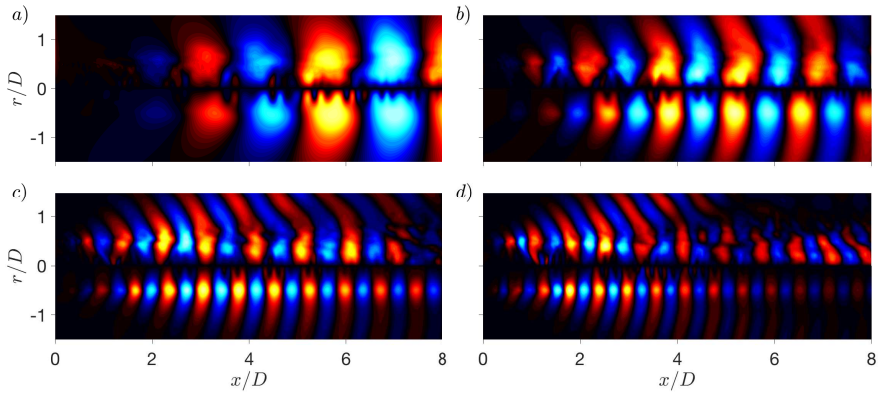}
\caption{Comparison between leading SPOD modes from LES (top) and the flow structure predicted by PFE (bottom) using $N=4$ and several frequencies. Real part of the modes are shown for $St=0.19$(a), $0.4$(b), $0.74$(c) and $0.91$(d). Only positive wavenumbers are allowed in the SPOD, and the PFE reconstruction is performed using non-negative components.}
\label{fig:PFELEScompreal_filt}
\end{figure}

Figures \ref{fig:PFELEScompabs_filt} and \ref{fig:PFELEScompreal_filt} show the comparison between filtered SPOD and non-negative components of PFE. Compared to figures \ref{fig:PFELEScompabs} and \ref{fig:PFELEScompreal}, the modulation of the SPOD modes seems consistently weaker for all frequencies, suggesting that the modulation relative to negative wavenumbers is stronger than its positive counterpart, or that the presence of upstream waves is responsible for a substantial part of the observed modulation; at this point, it is impossible to separate these two effects. PFE modes are also less strongly modulated, and most of the observed oscillations are concentrated around the centreline in figure \ref{fig:PFELEScompabs}. The filter is shown to improve the overall agreement, which can also be seen in the radial structure and wavelengths of the wavepacket, shown in figure \ref{fig:PFELEScompreal_filt}.

\begin{figure}
\centering
\includegraphics[clip=true, trim= 0 0 0 0, width=\textwidth]{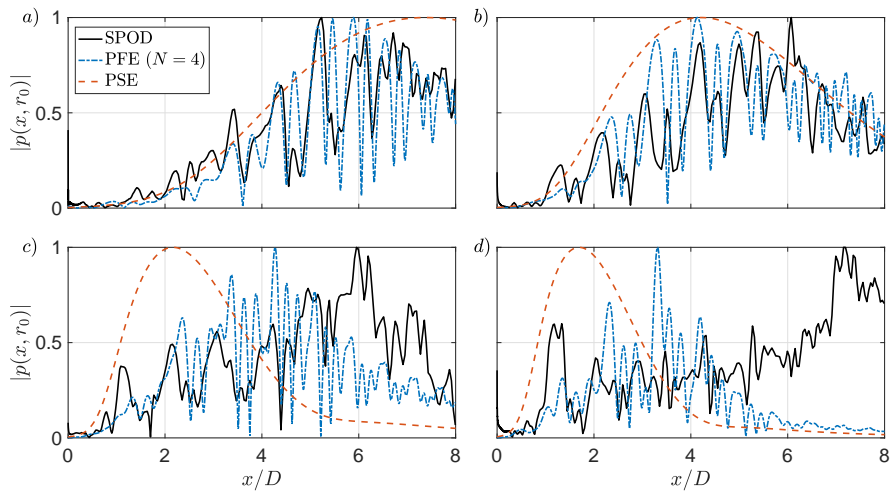}
\caption{Absolute value of leading SPOD modes from LES compared to PFE ($N=4$) and PSE results for several frequencies. Magnitude of the modes extracted at $r/D=0.1$ for $St=0.19$(a), $0.4$(b), $0.74$(c) and $0.91$(d). Only positive wavenumbers are allowed in the SPOD, and the PFE reconstruction is performed using non-negative components. All curves are normalised by their maximum.}
\label{fig:PFELEScompr01_filt}
\end{figure}

The amplitude modulations are compared for $r/D=0.1$ in figure \ref{fig:PFELEScompr01_filt}. Without the presence of the negative component of the modulation, the agreement between PFE and filtered SPOD improves substantially, especially at the early stations of the jet. Due to the presence of the second harmonic, the model is able to predict the position of both medium and large peaks, as smaller oscillations are associated with higher-order modulation components. Considering the limitations of the model highlighted in the previous section, this level of agreement suggests that appropriate filtering may permit the use of PFE even in cases where it would otherwise perform poorly.

\section{Conclusions}
\label{sec:concl}

In this work, we propose a formulation to predict modulated coherent structures in shock-containing jets. First, the shock-cell structure for a given flow condition is approximated using PSE applied to a modelled ideally expanded jet. This solution is then used to define a new shock-containing mean flow, around which the Navier-Stokes equations are linearised. By means of a Floquet ansatz, a new set of equations is obtained, in which the positive and negative components of the modulation (related to the local shock-cell wavenumber) are considered in the response vector. The final equations have the same shape as in PSE and are solved in the same fashion, rendering the name parabolised Floquet equations \citep{ran2019pfe}. We show that the formulation is more restrictive than PSE due to the repetition of the modes in the eigenspectrum of the linear operators (as in \cite{nogueira2021splsa}), which may also require substantially larger spatial steps for the march, depending on the flow conditions and the frequency of analysis.

The formulation is applied to two cases. The first considers an underexpanded jet at low supersonic Mach number, modelled by a canonical slowly-diverging flow. The solution procedure is shown in detail for two frequencies, including the definition of the initial solution to be marched downstream. The modulation characteristics of the resulting axisymmetric wavepacket are explored, and the initial results suggest that the modulation can be very sensitive to the different modes supported by the flow at a given frequency. Overall, some characteristics of the flow structures obtained using PFE are similar to results in the literature, which provided a first validation of the method. After that, the same methodology was applied to predict the most energetic (non-screeching) helical structures from an LES of an overexpanded jet. For that, a slightly more complex mean flow model was chosen to provide a better comparison with the LES data. Using such a model, PSE was able to predict the overall behaviour of the shock-cell structure, especially its leading wavenumber and the shock spacing variation further downstream. Using this solution, PFE modes were computed and compared to pressure SPOD modes for azimuthal wavenumber $m=1$. An overall good agreement was obtained in the predictions, especially closer to the centreline of the jet, suggesting that the method is able to capture the underlying physical mechanisms for the generation of these structures: the extraction of energy from the mean flow by the Kelvin-Helmholtz mode, and a redistribution of energy to modulation wavenumbers due to the interaction with the shock-cell structure, as predicted by \cite{TamTanna1982}. The model outperforms the classic PSE method, providing a good first approximation of the wavepackets supported by the flow, despite its several limitations. No experimental input is needed for the predictions, but a representative ideally expanded mean flow must be provided. While zero-frequency PSE is used to predict the overall behaviour of the shock-cell structure, a decomposition of a experimentally obtained shock-containing mean flow following the same structure as in the present case could also be used as input in the analysis.

The present framework opens new avenues for modelling and control of shock-containing jets. The modulated wavepacket model may be used for BBSAN \citep{wong_jordan_maia_cavalieri_kirby_fava_edgington-mitchell_2021} and installed jet noise \citep{Nogueira2017} predictions for generic flow conditions. It can also be used to make real-time predictions that could be applied in designing control strategies for noise reduction \citep{sasaki_piantanida_cavalieri_jordan_2017}. Furthermore, the overall Floquet formulation could also be used to build other prediction models, such as the one-way Navier-Stokes equations \citep{TOWNE2015844}. The explicit filtering of upstream waves in the one-way method has the potential to solve most of the issues of the present formulation, allowing for a clear separation between upstream waves and negative wavenumber modulation.  \\

\noindent \textbf{Acknowledgements.} This work was supported by the Australian Research Council through the Discovery Project scheme: DP190102220. We thank Guillaume Br\`es for generously providing the overexpanded jet LES database used in the present manuscript, and also Kenzo Sasaki for providing a first version of his PSE code.  \\

\noindent \textbf{Declaration of Interests.} The authors report no conflict of interest.

\appendix
\section{Linear operators for the parabolised Floquet equations}
\label{app:linops}

We can exemplify how the PFE solution is obtained by working with the continuity equation, which can be written as

\begin{equation}
    -\ii \omega \nu + \bar{u}\frac{\partial \nu}{\partial x} + u \frac{\partial \bar{\nu}}{\partial x} + v \frac{\partial \bar{\nu}}{\partial r} - \nu \frac{\partial \bar{u}}{\partial x} - \bar{\nu}\left[ \frac{\partial u}{\partial x} + \frac{\partial v}{\partial r} + \frac{v}{r} + \frac{\ii m}{r} w \right] =0.
    \label{eqn:Continuity1}
\end{equation}

Inserting (\ref{eqn:solPSE_shocks}) (using $N=1$) into (\ref{eqn:Continuity1}) and ignoring all the terms in $\Gamma_s^{\pm 2}$, we can write the continuity equation as a function of the different modulation components. In other formulations, such as non-linear PSE (or even SPLSA \citep{nogueira2021splsa}), the explicit dependence on a single wavenumber $\alpha_s$ allows us to write three different equations from the equation above: one dependent on $\Gamma_s^0=1$, one on $\Gamma_s$ and another one on $\Gamma_s^*=\Gamma_s^{-1}$. The same argument is used here, by considering the collection to be performed at each streamwise station (since the shock-cell wavenumber is slowly varying, this approximation should still be representative of the phenomenon). Using that assumption, we can separate the evolution of disturbances related to the different powers of $\Gamma_s$. Namely, for $\Gamma_s^0$ we can write

\begin{equation}
\begin{split}
    \left[ -\ii \omega \hat{\nu}_{0} + \bar{u}_{0} \left( \frac{\partial \hat{\nu}_{0}}{\partial x} + \ii \alpha \hat{\nu}_{0} \right) + \hat{u}_{0} \frac{\partial \bar{\nu}_0}{\partial x} + \hat{v}_{0} \frac{\partial \bar{\nu}_{0}}{\partial r}+ \right.  \\
    \left. - \hat{\nu}_{0} \frac{\partial \bar{u}_0}{\partial x} - \bar{\nu}_{0}\left( \frac{\partial \hat{u}_{0}}{\partial x} + \ii \alpha \hat{u}_{0} \right) - \bar{\nu}_{0}\left( \frac{\partial }{\partial r} + \frac{1}{r}\right) \hat{v}_{0} - \bar{\nu}_{0}\frac{\ii m}{r} \hat{w}_{0} \right] +  \\
    +\left[ u_s^* \left( \frac{\partial \hat{\nu}_{+1}}{\partial x} + \ii (\alpha+\alpha_s) \hat{\nu}_{+1} \right) + \hat{u}_{+1}\left( \frac{\partial \nu_s^*}{\partial x} - \ii \alpha_s \nu_s^* \right) + \hat{v}_{+1} \frac{\partial \nu_s^*}{\partial r}+ \right.  \\
    \left. - \hat{\nu}_{+1} \left( \frac{\partial u_s^*}{\partial x} - \ii \alpha_s u_s^* \right) - \nu_{s}^*\left( \frac{\partial \hat{u}_{+1}}{\partial x} + \ii (\alpha+\alpha_s )\hat{u}_{+1} \right) - \nu_{s}^*\left( \frac{\partial }{\partial r} + \frac{1}{r}\right) \hat{v}_{+1} - \nu_s^*\frac{\ii m}{r} \hat{w}_{+1} \right] +  \\
    +\left[ u_s \left( \frac{\partial \hat{\nu}_{-1}}{\partial x} + \ii (\alpha-\alpha_s) \hat{\nu}_{-1} \right) + \hat{u}_{-1}\left( \frac{\partial \nu_s}{\partial x} + \ii \alpha_s \nu_s \right) + \hat{v}_{-1} \frac{\partial \nu_s}{\partial r}+ \right.  \\
    \left. - \hat{\nu}_{-1} \left( \frac{\partial u_s}{\partial x} + \ii \alpha_s u_s \right) - \nu_{s}\left( \frac{\partial \hat{u}_{-1}}{\partial x} + \ii (\alpha-\alpha_s )\hat{u}_{-1} \right) - \nu_{s}\left( \frac{\partial }{\partial r} + \frac{1}{r}\right) \hat{v}_{-1} - \nu_s\frac{\ii m}{r} \hat{w}_{-1} \right]=0, 
    \end{split}
    \label{eqn:Continuity_Gamma0}
\end{equation}

\noindent for $\Gamma_s$

\begin{equation}
\begin{split}
    \left[ -\ii \omega \hat{\nu}_{+1} + \bar{u}_{0} \left( \frac{\partial \hat{\nu}_{+1}}{\partial x} + \ii (\alpha+\alpha_s)\hat{\nu}_{+1} \right) + \hat{u}_{+1} \frac{\partial \bar{\nu}_0}{\partial x} + \hat{v}_{+1} \frac{\partial \bar{\nu}_{0}}{\partial r}+ \right.  \\
    \left. - \hat{\nu}_{+1} \frac{\partial \bar{u}_0}{\partial x} - \bar{\nu}_{0}\left( \frac{\partial \hat{u}_{+1}}{\partial x} + \ii (\alpha + \alpha_s)\hat{u}_{+1} \right) - \bar{\nu}_{0}\left( \frac{\partial }{\partial r} + \frac{1}{r}\right) \hat{v}_{+1} - \bar{\nu}_{0}\frac{\ii m}{r} \hat{w}_{+1} \right] +  \\
    +\left[ u_s \left( \frac{\partial \hat{\nu}_{0}}{\partial x} + \ii \alpha \hat{\nu}_{0} \right) + \hat{u}_{0}\left( \frac{\partial \nu_s}{\partial x} + \ii \alpha_s \nu_s \right) + \hat{v}_{0} \frac{\partial \nu_s}{\partial r}+ \right.  \\
    \left. - \hat{\nu}_{0} \left( \frac{\partial u_s}{\partial x} + \ii \alpha_s u_s \right) - \nu_{s}\left( \frac{\partial \hat{u}_{0}}{\partial x} + \ii \alpha\hat{u}_{0} \right) - \nu_{s}\left( \frac{\partial }{\partial r} + \frac{1}{r}\right) \hat{v}_{0} - \nu_s\frac{\ii m}{r} \hat{w}_{0} \right]=0, 
    \end{split}
    \label{eqn:Continuity_Gammas}
\end{equation}

\noindent and equivalently for $\Gamma_s^*$.

The symmetry of the equations above becomes apparent by comparing the terms inside square brackets with (\ref{eqn:Continuity1}), which helps us to build the new equivalent system. By applying the same process to the Navier-Stokes and energy equations (and assuming $Re\to\infty$, for simplicity), we obtain

\begin{eqnarray}
    -\ii \omega 
    \begin{bmatrix}
    \mathbf{\hat{q}_{-1}} \\[0.5em]
    \mathbf{\hat{q}_{0}} \\[0.5em]
    \mathbf{\hat{q}_{+1}} 
    \end{bmatrix}
    +
    \begin{bmatrix}
    \mathbf{L_0} & \mathbf{L_{s^*}} & O \\[0.5em]
    \mathbf{L_{s}} & \mathbf{L_0} & \mathbf{L_{s^*}} \\[0.5em]
    O & \mathbf{L_{s}} & \mathbf{L_0}
    \end{bmatrix}
    \begin{bmatrix}
    \mathbf{\hat{q}_{-1}} \\[0.5em]
    \mathbf{\hat{q}_{0}} \\[0.5em]
    \mathbf{\hat{q}_{+1}} 
    \end{bmatrix}
    +
    \begin{bmatrix}
    \mathbf{B_0} & \mathbf{B_{s^*}} & O \\[0.5em]
    \mathbf{B_{s}} & \mathbf{B_0} & \mathbf{B_{s^*}} \\[0.5em]
    O & \mathbf{B_{s}} & \mathbf{B_0}
    \end{bmatrix}
    \frac{\partial}{\partial x}
    \begin{bmatrix}
    \mathbf{\hat{q}_{-1}} \\[0.5em]
    \mathbf{\hat{q}_{0}} \\[0.5em]
    \mathbf{\hat{q}_{+1}} 
    \end{bmatrix} + \nonumber \\
    + \ii \alpha
    \begin{bmatrix}
    \mathbf{B_0} & \mathbf{B_{s^*}} & O \\[0.5em]
    \mathbf{B_{s}} & \mathbf{B_0} & \mathbf{B_{s^*}} \\[0.5em]
    O & \mathbf{B_{s}} & \mathbf{B_0}
    \end{bmatrix}
    \begin{bmatrix}
    \mathbf{\hat{q}_{-1}} \\[0.5em]
    \mathbf{\hat{q}_{0}} \\[0.5em]
    \mathbf{\hat{q}_{+1}} 
    \end{bmatrix}
    + \ii \alpha_s
    \begin{bmatrix}
    -\mathbf{B_0} & O & O \\[0.5em]
    -\mathbf{B_{s}} & O & \mathbf{B_{s^*}} \\[0.5em]
    O & O & \mathbf{B_0}
    \end{bmatrix}
    \begin{bmatrix}
    \mathbf{\hat{q}_{-1}} \\[0.5em]
    \mathbf{\hat{q}_{0}} \\[0.5em]
    \mathbf{\hat{q}_{+1}} 
    \end{bmatrix}=0,
\label{eqn:FullSystem}
\end{eqnarray}

\noindent which is the explicit form of system (\ref{eqn:ShockPSE}) for $N=1$, with

\begin{eqnarray}
    \mathbf{L_t}=
    \begin{bmatrix}
    \mathbf{L_0} & \mathbf{L_{s^*}} & O \\[0.5em]
    \mathbf{L_{s}} & \mathbf{L_0} & \mathbf{L_{s^*}} \\[0.5em]
    O & \mathbf{L_{s}} & \mathbf{L_0}
    \end{bmatrix}, \
    \mathbf{B_t}=
    \begin{bmatrix}
    \mathbf{B_0} & \mathbf{B_{s^*}} & O \\[0.5em]
    \mathbf{B_{s}} & \mathbf{B_0} & \mathbf{B_{s^*}} \\[0.5em]
    O & \mathbf{B_{s}} & \mathbf{B_0}
    \end{bmatrix}, \
    \mathbf{B_{st}}
    \begin{bmatrix}
    -\mathbf{B_0} & O & O \\[0.5em]
    -\mathbf{B_{s}} & O & \mathbf{B_{s^*}} \\[0.5em]
    O & O & \mathbf{B_0}
    \end{bmatrix}.
\end{eqnarray}

The operators $\mathbf{L_0}$, $\mathbf{L_{s}}$, $\mathbf{L_{s^*}}$, $\mathbf{B_0}$, $\mathbf{B_{s}}$ and $\mathbf{B_{s^*}}$ are given by

\begin{eqnarray}
\mathbf{L_0}=
    \setlength{\arraycolsep}{4pt}
    \renewcommand{\arraystretch}{1.3}
    \left[
    \begin{array}{ccccc}
    - \partial_x \bar{u}_{0}  &  \partial_x \bar{\nu}_0 &  \partial_r \bar{\nu}_0-\bar{\nu}_0(D_r+\frac{1}{r})  &  -\ii m \frac{\bar{\nu}_0}{r}  &  0  \\
    \partial_x \bar{p}_0 & \partial_x \bar{u}_{0}  &  \partial_r \bar{u}_{0}  &  0  &  0  \\
    \partial_r \bar{p}_0  &  0  & 0 &  0  &  \bar{\nu}_0 D_r   \\
    0  &  0  &  0 &  0 & \ii m \frac{\bar{\nu}_0}{r}   \\
    0  &  \partial_x \bar{p}_0  &  \partial_r \bar{p}_0 + \gamma \bar{p}_0 D_r + \gamma \frac{\bar{p}_0}{r}  &  \ii m \gamma \frac{\bar{p}_0}{r} & \gamma \partial_x \bar{u}_{0}  \\
    \end{array}  \right], \nonumber \\
    \
    \label{eqn:ap.L0}
\end{eqnarray}

\begin{eqnarray}
\mathbf{L_s}=
    \setlength{\arraycolsep}{4pt}
    \renewcommand{\arraystretch}{1.3}
    \left[
    \begin{array}{ccccc}
    - (\partial_x u_s + \ii \alpha_s u_{s})  &  \partial_x \nu_s + \ii \alpha_s \nu_{s} &  \partial_r \nu_s-\nu_s(D_r+\frac{1}{r})  &  -\ii m \frac{\nu_s}{r}  &  0  \\
    \partial_x p_s + \ii \alpha_s p_{s} & \partial_x u_{s}+\ii \alpha_s u_{s}  &  \partial_r u_{s}  &  0  &  0  \\
    \partial_r p_s  &  0  & 0 &  0  &  \nu_s D_r   \\
    0  &  0  &  0 &  0 & \ii m \frac{\nu_s}{r}   \\
    0  &  \partial_x p_s + \ii \alpha_s p_{s}  &  \partial_r p_s + \gamma p_s D_r + \gamma \frac{p_s}{r}  &  \ii m \gamma \frac{p_s}{r} & \gamma (\partial_x u_{s} + \ii \alpha_s u_{s})  \\
    \end{array}  \right], \nonumber \\
    \
    \label{eqn:ap.Ls}
\end{eqnarray}

\begin{eqnarray}
\mathbf{L_{s^*}}=
    \setlength{\arraycolsep}{4pt}
    \renewcommand{\arraystretch}{1.3}
    \left[
    \begin{array}{ccccc}
    - (\partial_x u_s^* - \ii \alpha_s u_{s}^*)  &  \partial_x \nu_s^* - \ii \alpha_s \nu_{s}^* &  \partial_r \nu_s^*-\nu_s^*(D_r+\frac{1}{r})  &  -\ii m \frac{\nu_s^*}{r}  &  0  \\
    \partial_x p_s^* - \ii \alpha_s p_{s}^* & \partial_x u_{s}^*-\ii \alpha_s u_{s}^*  &  \partial_r u_{s}^*  &  0  &  0  \\
    \partial_r p_s^*  &  0  & 0 &  0  &  \nu_s^* D_r   \\
    0  &  0  &  0 &  0 & \ii m \frac{\nu_s^*}{r}   \\
    0  &  \partial_x p_s^* - \ii \alpha_s p_{s}^*  &  \partial_r p_s^* + \gamma p_s^* D_r + \gamma \frac{p_s^*}{r}  &  \ii m \gamma \frac{p_s^*}{r} & \gamma (\partial_x u_{s}^* - \ii \alpha_s u_{s}^*)  \\
    \end{array}  \right], \nonumber \\
    \
    \label{eqn:ap.Lsc}
\end{eqnarray}

\begin{eqnarray}
\mathbf{B_0}=
    \setlength{\arraycolsep}{5pt}
    \renewcommand{\arraystretch}{1.3}
    \left[
    \begin{array}{ccccc}
    \bar{u}_{0} &  -\bar{\nu}_0  &  0  &  0  &  0  \\
    0  &  \bar{u}_{0}  &  0  &  0  &  \bar{\nu}_0  \\
    0  &  0  &  \bar{u}_{0} &  0  &  0   \\
    0  &  0  &  0 &  \bar{u}_{0}  & 0   \\
    0  & \gamma \bar{p}_{0}  &  0  &  0  &  \bar{u}_{0}  \\
    \end{array}  \right],
    \label{eqn:ap.B0}
\end{eqnarray}

\begin{eqnarray}
\mathbf{B_s}=
    \setlength{\arraycolsep}{5pt}
    \renewcommand{\arraystretch}{1.3}
    \left[
    \begin{array}{ccccc}
    u_{s} &  -\nu_s  &  0  &  0  &  0  \\
    0  &  u_{s}  &  0  &  0  &  \nu_s  \\
    0  &  0  &  u_{s} &  0  &  0   \\
    0  &  0  &  0 &  u_{s}  & 0   \\
    0  & \gamma p_{s}  &  0  &  0  &  u_{s}  \\
    \end{array}  \right],
    \label{eqn:ap.Bs}
\end{eqnarray}

\begin{eqnarray}
\mathbf{B_{s^*}}=
    \setlength{\arraycolsep}{5pt}
    \renewcommand{\arraystretch}{1.3}
    \left[
    \begin{array}{ccccc}
    u_{s}^* &  -\nu_s^*  &  0  &  0  &  0  \\
    0  &  u_{s}^*  &  0  &  0  &  \nu_s^*  \\
    0  &  0  &  u_{s}^* &  0  &  0   \\
    0  &  0  &  0 &  u_{s}^*  & 0   \\
    0  & \gamma p_{s}^*  &  0  &  0  &  u_{s}^*  \\
    \end{array}  \right],
    \label{eqn:ap.Bsc}
\end{eqnarray}

\noindent where $D_r$ is the radial differential operator, $\partial_{x,r}$ indicates streamwise and radial derivatives of the mean flow quantities (including the shock-cell components) and $\gamma$ is the specific heat ratio.

\section{PSE results for shock-containing mean flow}
\label{app:PSEshock}

\begin{figure}
\centering
\includegraphics[clip=true, trim= 0 0 0 0, width=\textwidth]{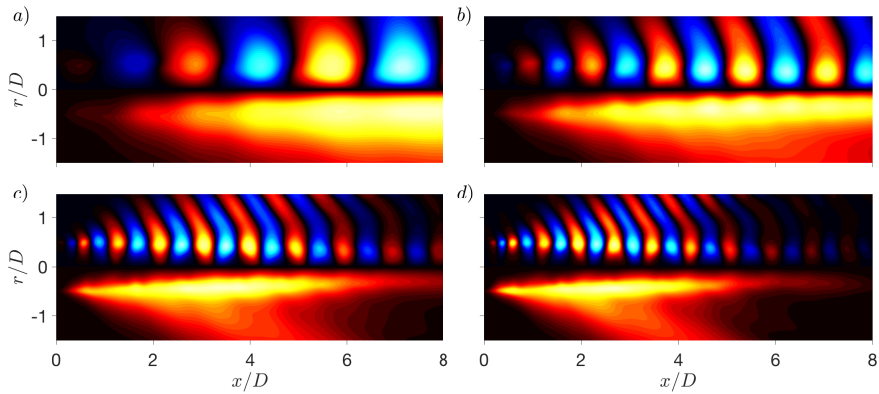}
\caption{Pressure fields predicted by PSE using the LES mean flow. Both real part (top) and absolute value (bottom) are shown for $St=0.19$(a), $0.4$(b), $0.74$(c) and $0.91$(d).}
\label{fig:PSE_shocks_new}
\end{figure}

Application of PSE requires a slowly varying mean flow and that the resulting functions $\mathbf{\hat{q}}$ and $\alpha$ are also slowly varying. While these conditions may be satisfied in perfectly expanded jets, shock-containing jet are usually considered to break these hypotheses a priori. The fast streamwise variation of these flows prohibits the coarse spatial marching of PSE, and results are more subject to numerical instability due to the sharp radial variations of the mean velocity caused by the shocks. Here, we ignore these assumptions and evaluate the PSE solution when a shock-containing mean flow is given as input. Sample pressure fields for the same frequencies studied in this work are shown in figure \ref{fig:PSE_shocks_new}. Interestingly, some modulation is observed in these wavepackets, mainly due to small local variations in the wavenumber $\alpha$. Still, since PSE only supports a single peak wavenumber (and small streamwise variations of it), this modulation is not comparable with the SPOD results, which are severely affected by the shock-cells. Overall, these results bear more resemblance to the shock-free PSE, with smooth variations of the resulting modes in the streamwise direction. That said, figure \ref{fig:PSE_shocks_new} shows that PSE captures the near-field sound generated by the wavepackets at these frequencies. As no quasi-periodicity is considered in this formulation, the distance between the KH and upstream modes in the eigenspectrum for higher frequencies is large, allowing for smaller spatial steps and, as a result, for a better representation of the sound generation mechanism.

\bibliographystyle{jfm}
\bibliography{ref.bib}

\end{document}